\documentclass[epsfig,amsmath,amssymb,twocolumn, usenatbib,a4paper]{mn2e} 

\usepackage{graphicx}
\usepackage{bm}
\usepackage{epsfig}
\usepackage{pifont} 
\usepackage{txfonts}
\usepackage{times}
\usepackage{yfonts}
\usepackage{soul}
\usepackage{ulem}

\fontencoding{T1}

\def\be{\begin{equation}}
\def\ee{\end{equation}}
\def\beq{\begin{equation}}
\def\eeq{\end{equation}}

\def\ba{\begin{eqnarray}}
\def\ea{\end{eqnarray}}

\newcommand{\LCDM}{$ \Lambda $CDM}

\title{The significance of the integrated Sachs-Wolfe effect revisited} 

\author[T.~Giannantonio, R.~Crittenden, R.~Nichol and A.~J.~Ross]{Tommaso Giannantonio$^{1,2}$\thanks{E-mail: tommaso.giannantonio at usm.lmu.de},
Robert Crittenden$^{3}$,
Robert Nichol$^{3,4}$, Ashley J. Ross$^{3}$\\
$^1$Ludwig-Maximilians-Universit\"at M\"unchen, Universit\"ats-Sternwarte M\"unchen, Scheinerstr. 1, D-81679 M\"unchen, Germany\\
$^2$Excellence Cluster Universe, Technical University Munich, Boltzmannstr. 2, D-85748 Garching bei M\"unchen, Germany\\
$^{3}$Institute of Cosmology and Gravitation, University of Portsmouth, Portsmouth, PO1 3FX, UK\\
$^{4}$SEPnet, South East Physics Network\\
}

\begin{document}

\date{\today}

\pagerange{\pageref{firstpage}--\pageref{lastpage}} \pubyear{2012}

\maketitle

\label{firstpage}

\begin{abstract}
We revisit the state of the integrated Sachs-Wolfe (ISW) effect measurements in light of newly available data and address criticisms about the measurements which have recently been raised.
We update the data set previously assembled by \citet{Giannantonio:2008zi} to include new data releases for both the cosmic microwave background (CMB) and the large-scale structure (LSS) of the Universe. We find that our updated results are  consistent with previous measurements. By fitting a single template amplitude, we now obtain a combined significance of the ISW detection at the $4.4 \, \sigma$ level, which fluctuates by $\sim 0.4 \, \sigma$ when alternative data cuts and analysis assumptions are considered.
We also make new tests for systematic contaminations of the data, focusing in particular on the issues raised by \cite{2010MNRAS.402.2228S}. Amongst them, we address the rotation test, which aims at checking for possible systematics by correlating pairs of randomly rotated maps. We find results consistent with the expected data covariance, no evidence for enhanced correlation on any preferred axis of rotation, and therefore no indication of any additional systematic contamination. 
We publicly release the results, the covariance matrix, and the sky maps used to obtain them.

\end{abstract}


\section{Introduction}

Observational evidence indicates the expansion of the Universe is accelerating at
late times, which may be explained by a small cosmological constant, some
negative-pressure dark energy fluid \citep{Frieman:2008sn}, by modifications of the laws of
gravity \citep[see e.g.][]{2011arXiv1106.2476C}, or by some non-trivial distribution of the local
large-scale structure \citep[see e.g.][]{Dunsby:2010ts}.
Evidence for this acceleration is provided by multiple complementary probes, such as observations of
distant Type Ia supernovae \citep{Amanullah:2010vv,2010MNRAS.401.2331L,Lampeitl:2010zx},  cosmic microwave background (CMB) anisotropies \citep{2011ApJS..192...18K}, baryon acoustic
oscillations (BAO) \citep{Eisenstein:2005su,2010MNRAS.401.2148P}, clusters of galaxies \citep{Rozo:2009jj}, and the integrated
Sachs-Wolfe (ISW) effect \citep{Giannantonio:2008zi,Ho:2008bz}.

We shall focus here on the latter, which consists of small secondary
fluctuations in the CMB which are produced whenever 
gravitational potentials are evolving, as happens at late times in the case of the
Universe undergoing a transition to a curvature- or dark
energy-dominated phase \citep{Sachs:1967er}. If we assume a flat universe as supported by
the primary CMB data, then a detection of the ISW represents a
measurement of dark energy and its properties.
Unfortunately, the amplitude of the ISW signal is small compared with the
intrinsic CMB temperature anisotropies, contributing to the signal only 
at large scales. To overcome this, a
technique was introduced to extract the ISW signal by
cross-correlating the observed CMB with tracers of the local
large-scale structure (LSS) of the Universe, such as wide-area galaxy
catalogues \citep{Crittenden:1995ak}.
As we will review below, this method has been used to successfully detect this signature of
dark energy by many authors using several different LSS catalogues and the WMAP 
data of the CMB. 
More recently, multiple data sets were analysed jointly to maximise the
extracted signal  \citep{Ho:2008bz, Giannantonio:2008zi} (G08 herein), thus
detecting the ISW
signal at an overall significance of $ \sim 4 \, \sigma$, when fit a
single amplitude.

However, some concerns have been raised about these detections, notably by
\cite{2010MNRAS.402.2228S, 2010MNRAS.406...14F, 2010MNRAS.406....2F, 2010A&A...520A.101H, 2010A&A...513A...3L}. These concerns relate to three main areas: conflicting estimates of the statistical significance;  searches based on new photometric data sets; and the possibility of larger than expected systematic contaminations.

The purpose of this paper is twofold. First we update our analysis to include the latest  data of both the CMB and the large-scale structure, using the 7-year WMAP maps and the latest available releases of the Sloan Digital Sky Survey (SDSS); we also publicly release the results of our analysis  and our sky maps. Second, we  evaluate  the above-mentioned criticism and  re-assess the overall state of the ISW measurements, focusing in particular on addressing the concerns by \cite{2010MNRAS.402.2228S} (S10 herein).

The plan of the paper is as follows: after reviewing the analysis techniques and the current state of the
ISW measurements in Section \ref{sec:status}, we will describe our updated data set and the publicly released maps in Section~\ref{sec:newdata}.
We then move on to a discussion of systematic uncertainties in Section~\ref{sec:systematics}, where we address a particular type of systematic test which has been discussed in S10 (the rotation test), finding  results consistent with those expected given the covariance of the data. 
Further issues raised by S10 and other authors are addressed in Section
\ref{sec:discussion}, before we conclude in Section \ref{sec:concl}.

\section{The state of the ISW} \label{sec:status}

\subsection {Theory}
The ISW effect \citep{Sachs:1967er} is a secondary source of temperature anisotropy which is produced whenever the gravitational potentials $\Phi$ and $\Psi$ are evolving in time, generating temperature anisotropies of the form
\be \label{eq:ISWbasic}
\Theta_{\mathrm{ISW}} (\hat \mathbf{n}) = - \int e^{- \tau (z)} \, \left( \dot \Phi + \dot \Psi \right) [\eta, \hat \mathbf {n}(\eta_0 - \eta)] \, d \eta \, ,
\ee
where $\eta$ is conformal time, the dots represent conformal-time derivatives, $\tau$ is the optical depth, and  $e^{- \tau (z)}$ is the CMB photon visibility function.  In the best-fit cosmological model, consisting of cold dark matter and a cosmological constant (\LCDM), such anisotropies are generated at early times during the transition from radiation to matter domination and at late times, when dark energy begins to dominate, so these two contributions are known as the early and late ISW effects. 

The early effect is typically generated shortly after recombination, so it is peaked around $l \sim 100 $, and its contribution to the total CMB (which is small in the standard \LCDM~case) can be used to constrain the energy content of relativistic species, such as the number of neutrino species and their masses \citep{Ichikawa:2008pz}, the presence of hot-dark-matter candidates such as massive neutrinos and axions \citep{2010JCAP...08..001H}, and interacting dark energy models \citep{Valiviita:2009nu}.

We focus here on the late effect as a probe of dark energy; in this case, the amplitude of the perturbations is also small compared with the primary CMB, and the fluctuations are generated on the largest scales, meaning that the effect is well described by linear theory.  
However, there is 
a small additional contribution from the non-linear growth in clusters, known as the Rees-Sciama effect \citep{Rees:1968zz}.  For a review, see \citet{Cooray:2002ee}; in more recent work, \citet{Smith:2009pn} provide a comparison with $N$-body simulations and perturbation theory, and \citet{2010MNRAS.407..201C} give a comparison of linear and non-linear effects on the reconstructed ISW maps from ray tracing of CMB photons through $N$-body simulations.  Finally, \citet{2011MNRAS.416.1302S} quantify the parameter estimation bias due to this non-linear effect and show that it is small compared to the statistical uncertainty imparted by cosmic variance.

\subsubsection{Cross-correlations}
Despite the small amplitude of the late ISW anisotropies, they can be used to constrain dark energy, as their presence can be detected by cross-correlating the observed CMB with the local matter density, which traces the gravitational potential \citep{Crittenden:1995ak}.
The CMB temperature anisotropy $\Theta_{\mathrm{ISW}}(\hat \mathbf {n})$ is given by Eq.~(\ref{eq:ISWbasic}), and  the galaxy density contrast $\delta_g(\hat \mathbf {n})$ can be calculated as
\be
\delta_g(\hat \mathbf {n}) = \int b_g(\hat \mathbf {n},z) \, \varphi (z) \, \delta (\hat \mathbf {n},z) \, dz \, ,
\ee
where $\delta(\hat \mathbf{n},z)$ is the dark matter density perturbation (linear theory suffices as described above), $b_g$ the (linear) galactic bias and $\varphi(z)$ is the normalised visibility function of the chosen galaxy survey.  This enables us to calculate the cross power spectrum, 
\be
C_l^{Tg , \, \mathrm{ISW}} = \frac{2} {\pi} \int dk \, k^2 \, P(k) \, W_l^{T, \, \mathrm{ISW}}(k) \, W_l^g(k) \, ,
\ee
where $P(k)$ is the matter power spectrum (linear theory suffices as well), and the source terms are, if we consider only the ISW temperature anisotropies,
\ba
W_l^{T, \, \mathrm{ISW}}(k) &=& -\int dz \, e^{-\tau(z)} \frac {d}{dz} \left[\tilde \Phi ( {k}, z) + \tilde \Psi( {k}, z) \right] \, j_l[k \chi(z)]  \nonumber \\
W_l^g(k) &=& \int dz \, \tilde b_g(k, z) \, \varphi (z) \, \tilde \delta ( {k},z) \,  j_l[k \chi(z)]  \, ,
\ea
where the tilde denotes Fourier transformation and the $j_l$ are the spherical Bessel functions.
The auto-power spectra for the galaxies $C_l^{gg}$ and the CMB (either the ISW part only $ C_l^{TT, \, \mathrm{ISW}}$ or the full observable spectrum $ C_l^{TT, \, \mathrm{tot}} $) can also be calculated by using the relevant source terms accordingly.
In this work we calculate all the theoretical predictions implementing the above equations into a modified version of the \textsc{camb} integrator code~\citep{Lewis:1999bs},
without using the Limber approximation.

Notice that in the \LCDM~model and in most of its variants, as long as secondary Doppler effects due to reionisation can be neglected \citep{Giannantonio:2007za}, the only significant source of large-angle CMB-density correlation is the ISW effect: we will therefore use the simpler notation $C_l^{Tg}$ for the cross-correlations.

\begin{center}
\begin{figure}
\includegraphics[width=\linewidth, angle=0]{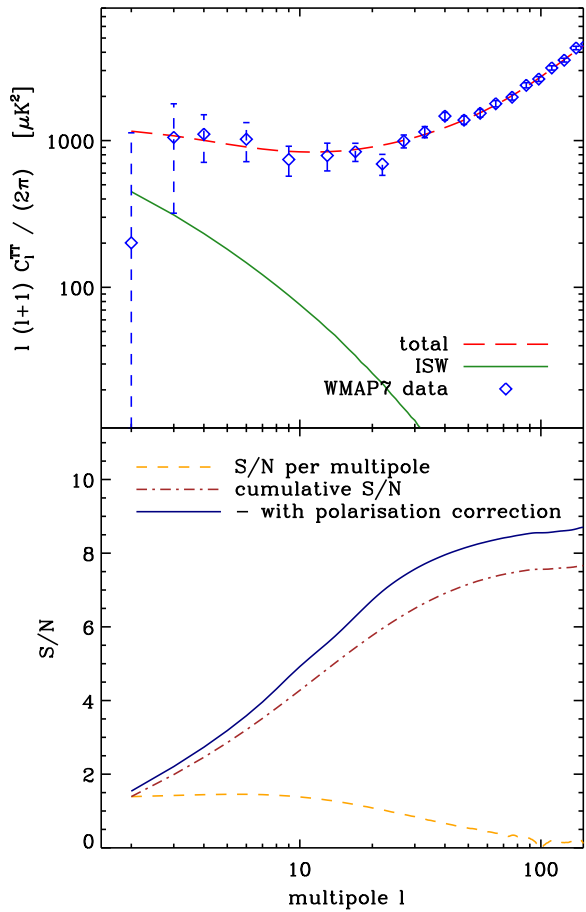}
\caption{Theoretical spectra and maximum signal-to-noise of the ISW detection for the current WMAP7 best-fit model \citep{2011ApJS..192...18K}. The top panel shows the angular temperature power spectrum of the ISW (green, solid) compared with the total CMB (red, dashed) and the data from WMAP7 (blue) \citep{Larson:2010gs}. The bottom panel shows the maximum signal-to-noise (per mode and cumulative) achievable given the same model as given by Eq.~(\ref{eq:snmax}) (brown, dot-dashed), as well as the improvement using CMB polarisation given by Eq.~(\ref{eq:snp}) (blue, solid).} 
\label{fig:SN}
\end{figure}
\end{center}

\subsubsection{Signal-to-noise}

The maximum signal-to-noise ($S/N$) ratio which is achievable for the ISW is limited by the amplitude of the primordial CMB perturbations.
For an idealised full-sky full-depth survey, one can write \citep{Crittenden:1995ak}
\be \label{eq:snmax}
\left( \frac{S}{N} \right)^2 \le \sum_{l} \, (2 l + 1) \cdot \frac {C_l^{TT, \, \mathrm{ISW}}}{C_l^{TT, \, \mathrm{tot}}} \, ;
\ee
for the current WMAP7 \LCDM~model \citep{2011ApJS..192...18K} described below in Section~\ref{sec:newdata}, this limit amounts to $S/N < 7.6$, as shown in Fig.~\ref{fig:SN}.
A more realistic estimation which takes into account the limitations of a galaxy survey, such as its redshift distribution, its shot noise due to finite surface density $n_s$ (in sr$^{-1}$) and its sky coverage $f_{\mathrm{sky}}$, is given by \citep[see e.g. G08;][]{2007MNRAS.381.1347C}:
\be  \label{eq:snreal}
\left( \frac{S}{N} \right)^2 \simeq f_{\mathrm{sky}} \sum_{l} \,  (2 l + 1)   \cdot  \frac{  \left(C_l^{Tg} \right)^2 }{ \left( C_l^{Tg} \right)^2 + C_l^{TT, \, \mathrm{tot}}  \,  \left( C_l^{gg} + 1 / n_s \right)  }  \, .
\ee
Applying this expression to detections coming
from current single galaxy catalogues typically yields only moderate significance  ($ 2 \lesssim S/N  \lesssim 3 $).

Early attempts were made to measure the two-point correlation between COBE CMB data and tracers of the LSS \citep{Boughn:2001zs}, but these were limited by the noise and resolution of the COBE data.  However, the ISW effect was soon detected using the WMAP data by many different groups exploiting a range of techniques;  the first detection by \citet{Boughn:2003yz} used the X-ray background from the HEAO satellite and radio-galaxies from the NVSS survey, and was followed quickly by many others as we review below.

\subsection{Single catalogue measurements}

The first ISW analyses were focused on detection at any significance, measuring the two-point correlations of the WMAP CMB data and  galaxy catalogues. This analysis can be performed equivalently in the real or harmonic spaces, using cross-correlation functions or cross power spectra.  While these approaches are formally equivalent, in practice some small differences can arise.  

\subsubsection{Cross-correlation functions}

In real space, the observable quantity is the cross-correlation function (CCF) between CMB temperature and galaxies defined as
\be
w^{Tg} (\vartheta) \equiv \langle \Theta (\hat \mathbf {n}_1) \, \delta_g (\hat \mathbf {n}_2) \rangle \, ,
\ee
where the average is carried out over all the pairs of directions in the sky lying at the given angular separation $\vartheta = |\hat \mathbf {n}_1 - \hat \mathbf {n}_2|$. 
This approach has the advantage of being computationally straightforward, since the sky masks are defined in real space and are easily treated; the main drawback is the high level of covariance between the measured data points.

Following the release of the WMAP data, several groups reported positive detections using a wide range of LSS catalogues: the first measurement by \citet{Boughn:2003yz} used a combination of NVSS radio galaxies and X-ray background data, and the NVSS result was independently confirmed by \citet{Nolta:2003uy}.  For optical surveys, indications were seen by \citet{Fosalba:2003iy} using the APM survey, and the evidence was improved by \citet{Fosalba:2003ge} and \citet{2003astro.ph..7335S} by using Luminous Red Galaxies (LRGs) from the SDSS.
\citet{2006MNRAS.372L..23C} detected the effect using the main SDSS galaxy sample, which while shallower, contains more galaxies than the LRG sample;  \cite{Rassat:2006kq} re-examined the relatively shallow 2MASS infrared survey, which was originally analysed with the power spectrum method by \citet{Afshordi:2003xu}.  In \citet{Giannantonio:2006du}, we reported the highest-redshift detection of the ISW with a catalogue of quasars from the SDSS, which was later reanalysed by \citet{2009JCAP...09..003X}. 

Most of these papers report a positive detection at low significance, typically between 2 and $ 3 \, \sigma$. One exception is the analysis using the 2MASS data where, though it favours the expected ISW signal, the evidence is very weak because the sample is too shallow for an appreciable ISW signal; it also is believed to have significant contamination from the Sunyaev-Zel'dovich (SZ) effect.

\subsubsection{Cross-power spectra}
We can also attempt to measure the angular power spectra of the cross-correlation directly. 
The main advantage of this approach is the relative decorrelation of the different modes, which makes the localisation of the signal on different scales more straightforward; the drawback is that the estimator of the correlation involves the inversion of a matrix whose dimensionality is the number of pixels over which the maps are projected ($N_{\mathrm{pix}}$), which is computationally challenging, especially in realistic cases where the geometry of the survey mask is complex. For these reasons, approximate methods are generally used \citep[for details, see e.g.][]{Padmanabhan:2002yv,Efstathiou:2003dj,Hirata:2004rp,Ho:2008bz,2012arXiv1203.3277S}, which still yield considerably lower correlation between modes when compared to cross-correlation measurements. 

In this way, positive detections were reported  by \citet{Afshordi:2003xu} using the 2MASS catalogue, who simultaneously fit a template for the SZ effect; this was recently revisited by \citet{2010MNRAS.406....2F} who found weaker evidence more consistent with \citet{Rassat:2006kq} and other analyses.  \citet{Padmanabhan:2004fy} applied the cross-spectrum technique to a SDSS LRG sample, and found consistent significance levels to the earlier work, at about $2 \, \sigma$.
More recently, \citet{2012MNRAS.422L..77G} measured the correlation between WMAP and the Wide-field Infrared Survey Explorer (WISE) survey, finding a high ISW signal at $> 3 \, \sigma$. While the WISE volume is $\sim 5$ times larger than 2MASS, and thus the expected signal is higher, such a high correlation is $\sim 2.2 \, \sigma$ higher than the \LCDM~expectations. Future analyses with the upcoming larger WISE data releases will help to clarify this issue.

 Given accurate covariance matrices, cross-correlation and cross-spectra measurements (using the same data), should yield identical results, as, in total, both measurements contain the same information.

\subsubsection{Other techniques}  
The ISW effect has also been seen using a method based on a wavelet decomposition by \citet{Vielva:2004zg, 2007MNRAS.376.1211M, 2008MNRAS.384.1289M, Pietrobon:2006gh}, who explored its dependence on different wavelet shapes and scales. While the significance level was sometimes reported to have been enhanced with this technique, the resulting constraints on cosmology were comparable with the previous two methods.

As the ISW is maximum on the largest scales, it is affected by the local variance, i.e. by the particular realisation of the matter distribution given the power spectrum; this may bias the results. For this reason, more advanced methods were developed to subtract the local variance, e.g. by \citet{2008A&A...490...15H, 2008MNRAS.391.1315F}. In the latter work, a Wiener filter reconstruction of the LSS was used as a template instead of the theoretical cross-correlation function; it was estimated that this method may increase the S/N ratio by 7\% in average.

It is also possible to reconstruct the ISW temperature maps. This was attempted by \citet{2008ISTSP...2..747B} using NVSS data and a Wiener filter method, and by \citet{Granett:2008dz} using SDSS data (see Section~\ref{sec:higher}).
Another optimised method has been later introduced by \citet{2011A&A...534A..51D}, based on the analysis of the temperature and density fields themselves rather than their spectra. A useful byproduct of this procedure is that a map of the ISW signal in the CMB is obtained. These authors also highlight the importance of separating the different statistical analyses, defining different procedures for testing the detection of a correlation in a model-independent way, measuring the confidence level based on a template, and comparing different models. This method was then validated with the 2MASS data, recovering a weak positive detection.

Another strategy to improve the signal-to-noise is to use the CMB polarisation information to reduce the primary CMB anisotropies, as proposed by \citet{2006eccc.confE...2C, 2009MNRAS.395.1837F, Liu:2010re}. Depending on the details of the method, different authors estimate a level of improvement in the significance of the ISW detection in the range between $5$ and $20\%$. 
In more detail, the correlation $C_l^{TE}$ between CMB temperature and polarisation ($E$-modes)  can be used  to reduce the amount of primary anisotropies in the total temperature spectrum. Assuming idealised data, the maximum signal-to-noise of Eq.~(\ref{eq:snmax}) is thus increased to
\be \label{eq:snp}
\left( \frac{S}{N} \right)^2 \le \sum_{l} \, (2 l + 1) \cdot \frac {C_l^{TT, \, \mathrm{ISW}}}{C_l^{TT, \, \mathrm{tot}} - \left. \left( C_l^{TE} \right)^2  \middle/ \, C_l^{EE} \right.} \, .
\ee
For the current WMAP7 \LCDM~model \citep{2011ApJS..192...18K}, we find that this improves the upper limit to $S/N < 8.7$, as shown in Fig.~\ref{fig:SN}.

Most of the approaches effectively measure a two-point statistic of the average correlations between CMB and LSS over the whole region covered by the surveys; however, with wavelets it is possible to try to localise sources of the ISW effect. This was more directly attempted by \citet{2008ApJ...683L..99G}, who identified massive superclusters and voids of galaxies in the SDSS LRG survey and their corresponding regions from WMAP were stacked to maximise the signal.
 A high significance detection was reported (at $4.4 \, \sigma$ from this LRG catalogue alone), although this result was strongly dependent on the number of superclusters and voids used. See below Section~\ref{sec:higher} for a more detailed discussion.

\subsection{Multiple catalogues measurements and their applications}

The significance of the ISW detections can be increased by combining measurements obtained with multiple catalogues, to improve from a simple detection to  parameter estimation and model comparison.
\citet{Gaztanaga:2004sk} made a first attempt at collecting all the existing measurements and used the resulting compilation to constrain cosmology; this was also extended by \citet{Corasaniti:2005pq}.

The difficulty with combining multiple measurements is achieving a reliable estimation of the covariance between them. It was proposed that a full tomographic analysis should be performed \citep{Pogosian:2005ez}, including all the signals, and their covariances, as a function of redshift. This was finally achieved independently by \citet{Ho:2008bz}, using the harmonic space estimator, and by \citet{Giannantonio:2008zi} in  real space, using five\footnote{In the analysis by \citet{Ho:2008bz} some of the catalogues where subdivided further into sub-samples.} and six galaxy catalogues respectively, summarising the state of the art in the field and  upgrading the significance to $3.7 \, \sigma$ and $4.5 \, \sigma$ respectively. These results have been used to test a variety of dark energy and modified gravity models \citep{Giannantonio:2008qr,2010JCAP...04..030G,Lombriser:2009xg,2010arXiv1003.3009L,Lombriser:2011tj,2009PhRvD..80l3516V,Serra:2009yp,Daniel:2009kr,Zhao:2010dz,Bertacca:2011in};
in the modified gravity case, the ISW provides a particularly useful constraint because it is sensitive to any non-trivial evolution of the gravitational potentials and the effective anisotropic stress.

\subsection{Potential issues}

Alongside these developments, some studies have questioned individual aspects  of the ISW measurements, raising some doubts about the significance of its detection.

In \cite{2010MNRAS.406...14F,2010MNRAS.406....2F}, the 2MASS-CMB cross-correlation was re-analysed, and it was found that there is little evidence for an ISW detection from this catalogue alone, which is in agreement with most previous literature.  But more worryingly, these authors also state that the ISW signal may remain undetected in 10\% of cases (see Section \ref{sec:FandP} below). 

In \cite{2010A&A...520A.101H}, the NVSS catalogue was re-considered, looking at its auto- and cross-correlation functions in both real and harmonic spaces. In both cases, some cross-correlation was seen, but the paper expressed concerns regarding a lower than expected signal on the largest scales and anomalous large-scale structure in the NVSS map. We discuss these issues in Section \ref{sec:HM}.

\citet{2010MNRAS.402.2228S} reanalysed some of the earlier ISW measurements and extended the analysis with three new LRG data sets, including a high-redshift sample developed using AAOmega spectra. The two catalogues at lower $z$ were found to be in general agreement with a positive ISW signal, although at lower significance than seen elsewhere in the literature, while the high-redshift AAOmega sample showed no significant correlation. These authors also discussed the effect of possible systematics, suggesting that there is evidence of strong residual systematics from the study of data generated by rotating the real maps.  We explore the rotation tests for our data  and for the S10 data in Section~\ref{sec:rotation}; we then discuss  the new data sets by S10 in Section \ref{sec:S09data}.

Finally, \citet{2010A&A...513A...3L} reviewed some of the correlation analyses, finding levels of signal comparable to previous measurements, but significantly higher levels of uncertainty. We discuss this further in Section \ref{sec:LC}.

\section{Updated data set} \label{sec:newdata}

We have updated our data compilation from G08 to include the latest available data for both the CMB and the LSS. All the data sets are pixellated in the Healpix scheme \citep{Gorski:2004by} at a resolution of $N_{\mathrm{side}} = 64$, corresponding to a pixel side of 0.9 degrees, as previously done in G08. We have checked that higher resolutions give consistent results. 

In the following analysis, unless otherwise stated, we assume a fiducial flat \LCDM~cosmology, which we here update to the latest best-fit model from WMAP7+BAO+$H_0$, defined by energy densities for baryons $\omega_b = 0.0226$, cold dark matter $\omega_c = 0.1123$, sound horizon at the last-scattering surface $ 100 \, \vartheta_* = 1.0389$, optical depth $\tau = 0.087$, spectral index and amplitude of primordial scalar perturbations $n_s = 0.963$, $A_s = 3.195$, referred to a pivot scale $k_\mathrm{pivot} = 0.002\, h/$Mpc \citep{Larson:2010gs,2011ApJS..192...18K}. 

We summarise in Table~\ref{tab:data} the most important properties of the data sets we use.

\begin {table*}
\begin {center}
\begin{tabular}{l | c | c | c | c | c | c} 
\hline
Catalogue    & band  &    $N$ after masking   & $f_{\mathrm{sky}} $  & $ n_s $ [sr$^{-1}$]  &  $\bar z$  &  $b$ \\ 
\hline
2MASS        & IR &    415,459  &  0.531  &  $6.23 \cdot 10^{4}$  &  0.086  &  1.3     \\
SDSS gal DR8 & Optical & 30,582,800  &  0.253  &  $9.60 \cdot 10^{6}$  &  0.31   &  1.2     \\
SDSS LRG DR7 & Optical &    918,731  &  0.181  &  $4.03 \cdot 10^{5}$  &  0.50   &  1.7      \\
\hline
NVSS         & Radio &  1,021,362  &  0.474  &  $1.72 \cdot 10^{5}$  &  1.05  &  1.8      \\ 
HEAO         & X  & N/A (flux)  &  0.275  &  N/A (flux)           &  0.90  &  1.0      \\
SDSS QSO DR6 & Optical &    502,565  &  0.168  &  $2.39 \cdot 10^{5}$  &  1.51   &  2.6      \\
\hline
\end{tabular}
\caption{Summary of the properties of the LSS catalogues used. We report the number of objects after masking $N$, the sky fraction $f_{\mathrm{sky}} $, the surface density of sources $n_s$, the median redshift of their distributions $\bar z$ and the galactic bias $b$ assumed constant needed to fit the auto-correlation function assuming the WMAP7 cosmology.}
\label{tab:data}
\end{center}
\end {table*}

\subsection {CMB data}
The original data set by G08 was obtained by analysing the maps from the third year of WMAP, and it was checked upon the release of the WMAP five-year data that it yielded  consistent results, as mentioned in Section IV.B of G08. 
Here we have updated the whole analysis by using the latest WMAP7 data \citep{Jarosik:2010iu}, which should give a more stable foreground subtraction and noise reduction due to the increased integration time.

As for the choice of frequency, we use the internal linear combination (ILC) map and we also use the most aggressive galaxy mask associated with it, again on the basis that this should include the best foreground subtraction.
We have checked that the signal does not change significantly between the different WMAP data releases, and that it is reasonably frequency-independent and close to the ILC result in the range of the WMAP bands Q, V, W. We discuss this further below.

\subsection {Main SDSS galaxy data} \label{sec:SDSS}
The main galaxy distribution from the SDSS has been extended from Data Release Six (DR6) to the final imaging SDSS-III (DR8) public data release \citep{Aihara:2011sj}. These galaxies have been selected using the same criteria as in G08, i.e. starting from the photo-z primary galaxy sample (\texttt{mode=1, type=3}), which contains 208 million objects, and then imposing a cut in redshift of $0.1 < z < 0.9$ and a cut in flux of $ 18 < r < 21 $, where $r$ is the $r$-band model magnitude corrected for extinction. Also, only objects with photo-z uncertainty of $\sigma_z(z) < 0.5 \, z$ were considered. This leaves us with $\sim 40$ million galaxies, with a redshift distribution centred around $z \simeq 0.3$.  As the distribution of the photo-z's can occasionally be inaccurate, for the calculation of the theoretical predictions we used a fit to a distribution function of the form introduced by \citet{1994MNRAS.270..245S} and given by
\be \label{eq:dndz}
\varphi(z) = \frac{1}{\Gamma \left(\frac {\alpha + 1} {\beta} \right)} \, \beta \, \frac {z^{\alpha}}{z_0^{\alpha+1}} \, \exp \left[ - \left( \frac{z}{z_0} \right)^{\beta} \right],
\ee
where the best-fit values of the parameters are $\alpha = 1.5$, $\beta = 2.3$, $z_0 = 0.34$; this fitted redshift distribution is similar to the DR6 result.

The mask was derived from a higher number density sample of the SDSS galaxies, selected with the weaker conditions $r < 22 $ and $ \Delta z < z $ (120 million objects), which was pixellated at a higher resolution ($N_{\mathrm{side}} = 512$) and finding the number of filled high-resolution sub-pixels in each low-resolution pixel. 
Each low-resolution pixel was then assigned a weight $f^g_i$ proportional to the fraction of high-resolution pixels which were filled.

{An additional subtlety here is to avoid biasing the mask due to the high-resolution pixels which are on the edge of the survey themselves. We have found that the count-in-cells distribution of the 120 million galaxies in the high-resolution pixels is well approximated by a log-normal distribution of median $\mu = \ln (110)$, and is even better fit by the gravitational quasi-equilibrium distribution (GQED), described by \citet{Yang:2010qs}.  By comparing the best-fitting GQED distribution with the data, we found that the survey's edges introduce an enhanced tail in the distribution at low occupation number which leads to a bias in the mask. For this reason, we remove from the mask all high-resolution pixels with $n<40$, since the best-fit GQED is nearly zero below this point.  We found that our results are not overly sensitive to reasonable differences in the value chosen for this threshold.}

We then mask the sky areas most affected by galactic extinction, dropping all pixels where the median extinction in the $r$ band is $A_r > 0.18$. {We have found that increasing this cut to the stricter level of $A_r > 0.16$ only changes the observed CCF by $5\%$.}
The unmasked survey area increased  from  7,771  pixels (or 16\% of the sky) for DR6  to  11,715 (or 24 \%) for DR8. The DR8 is the first data release to include a significant fraction of data from the Southern galactic hemisphere. We have checked that no significant difference appears when excluding data from the Southern hemisphere: the differences in the observed CCF are at the $10\%$ level.

We estimated the galactic bias by fitting the \LCDM~prediction to the observed auto-correlation function (ACF), and found a value $b = 1.2$ assuming a scale- and redshift-independent bias.

\subsection {Luminous Red Galaxies} \label{sec:LRG}

We update the catalogue to include the latest data release of the MegaZ LRGs by \citet{2011MNRAS.412.1669T}, which corresponds to the SDSS DR7, increasing our previous DR6 coverage by $10 \%$. We apply the completeness cut in the de-reddened deVacouleurs $i$ magnitude suggested by the authors ($i_\mathrm{deV} - A_i < 19.8$) and we limit the star-galaxy separation parameter to $\delta_{sg} > 0.2$. We finally apply the reddening mask and discard pixels of median extinction at $A_r > 0.18$ as in the main galaxy case. {As above, increasing this cut to the stricter level of $A_r > 0.16$ only changes the observed CCF by $5\%$.} The redshift distribution in this case peaks around $z = 0.5$ and is smooth, and we use it directly as done in G08. 
The mask for the LRGs is that provided by \citet{2011MNRAS.412.1669T}, with the addition of the aforementioned extinction mask.
The bias of this catalogue is found to be $b = 1.7 $, again by fitting to the measured ACF.

\subsection{QSO data}
For the quasars, no updated catalogue is yet publicly available, and here we use the same DR6 catalogue \citep{Richards:2008eq} as in G08, limiting ourselves to the cleaner subset of $UVX$-selected objects. To reduce the stellar contamination, which is more of an issue for quasars, we choose here a stricter extinction cut discarding pixels of $A_r > 0.14$. {Increasing this cut to the stricter level of $A_r > 0.12$  changes the observed CCF by less than $10\%$.} 

In this paper we  used Eq.~(\ref{eq:dndz}) to determine a fit to the QSO redshift distribution, for the same reasons described in Section~\ref{sec:SDSS}, and obtained  best-fit parameters of $\alpha = 2.0$, $\beta=1.5$, $z_0 = 1.06$, corresponding to a median $\bar z = 1.5$. This is changed from G08 where the visibility function was simply binned, and not smoothed, and so included irregular steps between adjacent bins. 

The linear bias parameter found from the QSO ACF is $b = 2.6 $, $10\%$ higher than that reported in G08. The amount of stellar contamination, as seen by comparing the large-scale power in the ACF with the ACF of a catalogue of stars from the SDSS, is consistent with a fraction $\kappa = 2 \%$, which is expected in these data \citep{Richards:2008eq}.

\begin{center}
\begin{figure}
\includegraphics[width=\linewidth, angle=0]{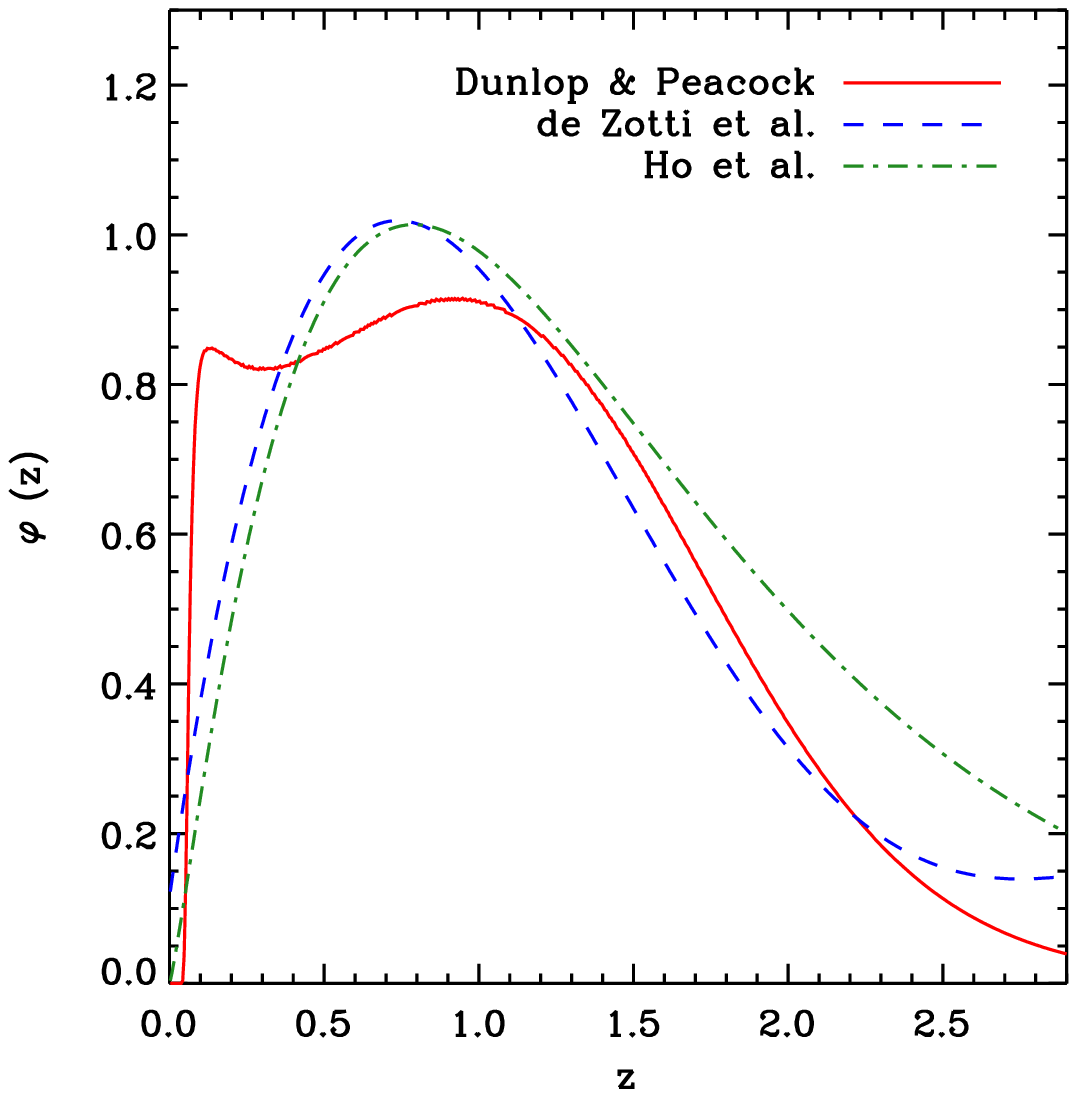}
\includegraphics[width=\linewidth, angle=0]{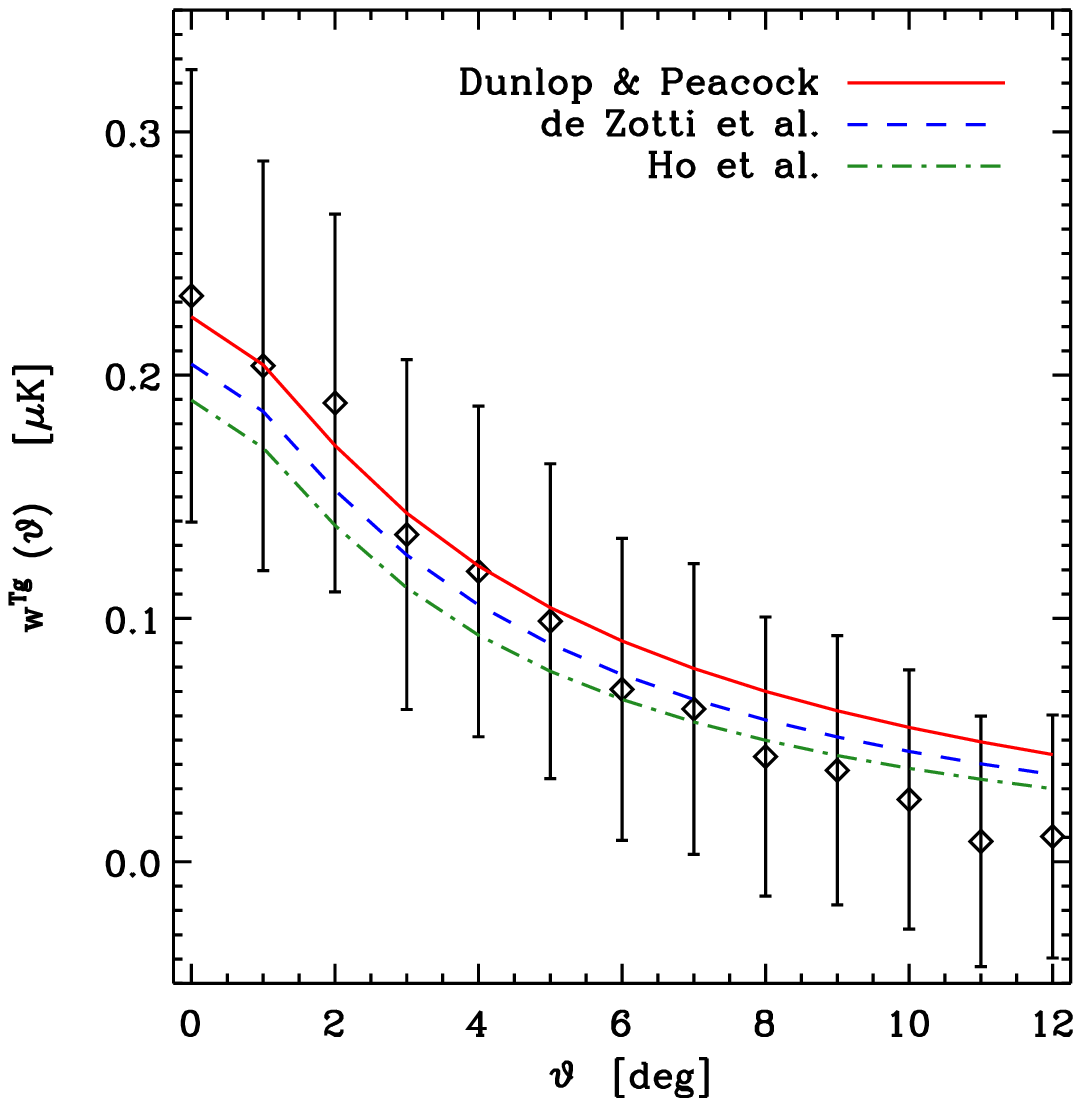}
\caption{The redshift distribution of NVSS and its effect on the ISW correlation. In the top panel we show three normalised selection functions which have been  used widely in the literature, and in the bottom panel the resulting theoretical  cross-correlation functions, compared with our  data, for the fiducial \LCDM~model. The bias is  constant, and  set to $b=1.8$ for the first two models as this gives  good agreement with the ACF. It is $b=1.98$ for the last model, as fitted by \citet{Ho:2008bz}.  We can see that the differences are small compared with the measurement error bars. We use the model by \citet{DeZotti:2009an} in the main analysis.}
\label{fig:nvssdndz}
\end{figure}
\end{center}

\subsection{Other data}
For the other surveys (2MASS, HEAO, and NVSS), we continue using the same maps as in the previous analysis of G08. However, we have improved the analysis in the following ways:

For the low redshifts probed by 2MASS, non-linear effects are large enough to significantly affect the zero-lag bin of the ACF; this was seen by 
comparing the linear power spectrum with the result obtained  by \citet{Smith:2006ne} at the scale of one degree. We have therefore dropped this bin which is significantly higher than the linear theory prediction, so that the linear bias has decreased to $b = 1.3 $ in this case.

For NVSS, a well-known issue is the uncertainty in the redshift distribution of the sources.  Significant changes to the redshift distribution can affect the cross-correlation template, and so impact the detection significance, and will also change the predicted   \LCDM~cosmology amplitudes.  Here we have compared the correlation functions obtained using the distribution function based on the models of \citet{Dunlop:1990kf} (used in G08), to the distribution from \citet{DeZotti:2009an}, who fitted the template based on radio galaxies with measured redshifts, and the distribution function introduced by \citet{Ho:2008bz}\footnote{Note the typo in Eq.~(33) of \citet{Ho:2008bz}, where the argument of the Gamma function should be $(\alpha+1)$ to ensure the stated normalisation of $\int f(z) dz = b_{\mathrm{eff}}$.}, who simultaneously fit the cross-correlation functions between NVSS and other galaxy catalogues. 
As seen in Fig.~\ref{fig:nvssdndz}, the theoretical predictions for these models are compatible within the expected measurement error bars. We find that the resulting significance of the NVSS ISW detection changes at most by 10\% when using each of the three models for the source distribution. In the following, we use the model by \citet{DeZotti:2009an}, as it is based on a subsample of galaxies of known redshifts.
 Further, we have included a better modelling of the shot noise in the ACF, due to the fact the maps were originally pixellated at a lower resolution. This primarily affects the ACF, and so the measurement of the bias;  we now find $b = 1.8 $, increased from $b = 1.5$ previously assumed in G08.

For the HEAO catalogue, we  have also included the same pixellation correction in the shot noise modelling, but as the instrumental beam is  much larger ($\vartheta_{\mathrm{FWHM}} = 3.04$ degrees), the resulting bias parameter remains $b = 1.0 $.

\subsection {Method}
We pixellate all these data on the sphere as described above, and
measure the two-point functions between them and the CMB, using the simple estimator 
\be
\hat w^{Tg} (\vartheta) = \frac{1}{N_{\vartheta}} \sum_{i,j}^{N_{\mathrm{pix}}} \frac{ \left( n_i - \bar n \right)}{\bar n} \, \left(T_j - \bar T \right) \, f^g_i \, f^T_j \, ,
\ee
where $n_i$ and $ T_i$ are the number of galaxies and the CMB temperature in a pixel of masked weights $f^g_i, f^T_j$ respectively, $\bar n, \bar T$ are the average number of galaxies per pixel and average temperature, and 
$N_{\vartheta} = \sum_{i,j} f^g_i f^T_j$ is  the weighted number of pairs at a given separation. Note that in our approach the CMB weights $f^T_j$ are simply taken to be either 0 or 1, depending on whether the pixel is masked or unmasked. When $f^g_i < 1$, which occurs mostly at the edges of the surveyed area, the number of galaxies in a pixel $n_i$ needs to be rescaled from the observed number $n_i^{\mathrm{obs}}$ as $n_i = n_i^{\mathrm{obs}}/f_i^g$, in order to keep the same mean density $\bar n$ over the whole map.
As in G08, we use 13 angular bins linearly spaced between 0 and 12 degrees.

We estimate the full covariance matrix $\mathcal{C}$ using the `MC2' Monte Carlo method described in G08; specifically, based on the fiducial flat \LCDM~cosmology, we generate Gaussian random maps of the CMB and of all the galaxy catalogues, using their known redshift distributions, number densities, and including all the expected correlations between the catalogues. We also add the expected level of Poisson noise based on the surface densities of each catalogue on top of all realisation of the Gaussian maps. For each of 5000 realisations of these maps, we measure the correlation functions, and calculate the covariance of them.  (See the Appendix of G08 for more details.) We have confirmed that 5000 realisations are enough for convergence of the signal-to-noise. As the cross-correlations are in agreement with the fiducial \LCDM~cosmology used in the mocks, we expect this modelling of the covariance should be reasonably accurate. See below Section~\ref{sec:comparecvm} for more details and a comparison with the analytic covariance.
Most catalogues are significantly covariant; the samples which are less covariant with the others are the LRGs and the QSOs, because of their unique redshift coverage, which is more peaked in the former case, and deeper in the latter.

We fit the amplitudes assuming the cross-correlation functions are Gaussianly distributed; this is approximate, as even if the maps themselves are Gaussian, as assumed in our Monte Carlos, two point statistics of those maps will not be Gaussian.  However, this appears to be a reasonable approximation for correlation functions (and particularly cross-correlations), where each bin represents an average of many products of pixels and the central limit theorem should apply.  This is confirmed in our Monte Carlos, where we find the skewness in the covariance to be relatively small, with dimensionless skewness measure of $0.1-0.2$.  However, it may be worth further investigating any residual bias which this level of non-Gaussianity might cause in future work.  Non-Gaussianity is likely to be a more significant concern for power spectrum estimators, particularly for auto-spectrum measurements which must be positive-definite. 

\begin{center}
\begin{figure*}
\includegraphics[width=.9\linewidth, angle=0]{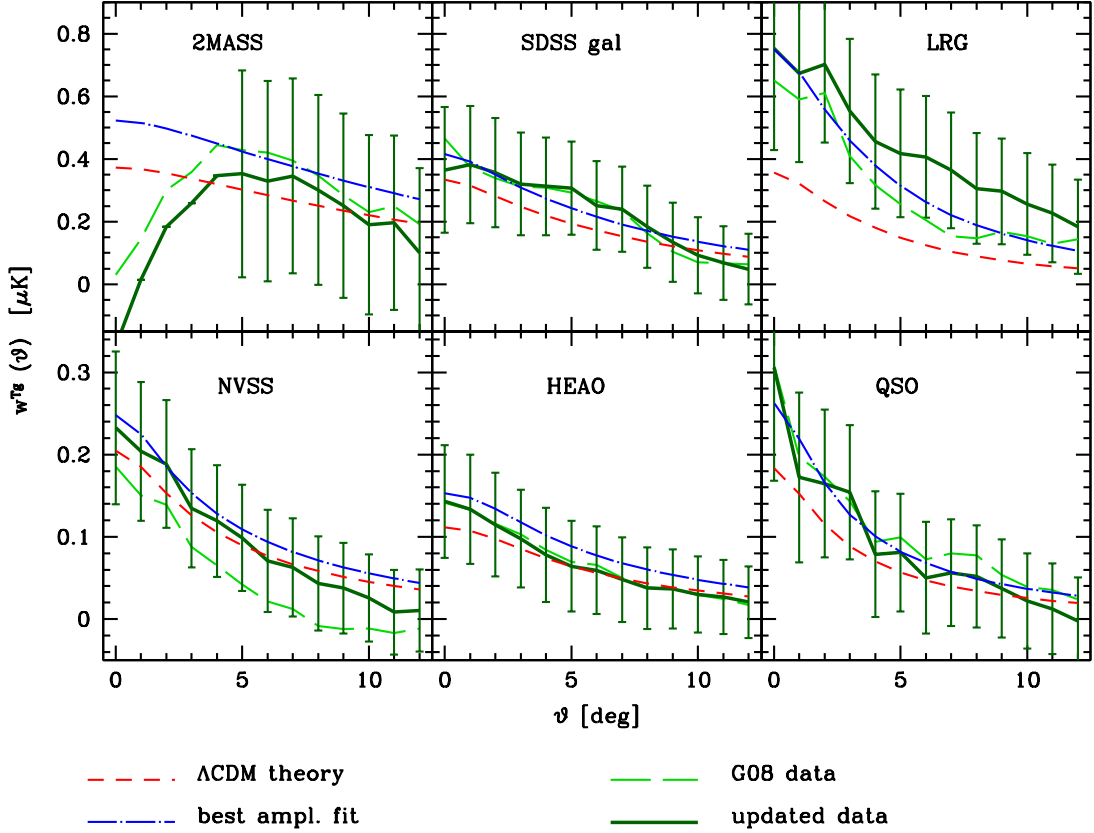}
\caption{Updated results of the cross-correlation of all the data sets with the WMAP7 ILC map. Most data (dark green, solid) are in good agreement with the theoretical predictions for a \LCDM~model (red, short-dashed), with the only exception of the LRGs which show an excess at  $>$1-$\sigma$ level.  The highly correlated error bars are from 5000 Monte Carlo mocks and are 1-$\sigma$, except for 2MASS where they are 0.5-$\sigma$ to improve readability of the plot. Further, the first five data points for 2MASS have been excluded due to potential contamination by the SZ effect. The light-green, long-dashed lines show the previously published data by G08, and the blue, dot-dashed lines are the best amplitude fits.}
\label{fig:wmap7}
\end{figure*}
\end{center}

\subsection {Results and public release}

The results of the new cross-correlation analysis are shown in Fig.~\ref{fig:wmap7}, and are in general agreement with G08 given the measurement errors.  We can see that all the measurements lie close to the \LCDM~prediction;  however the LRGs do show an excess signal at the $> 1 \, \sigma$ level. See Section~\ref{sec:systematics} for a discussion of possible systematic effects.

\begin{center}
\begin{figure}
\includegraphics[width=\linewidth, angle=0]{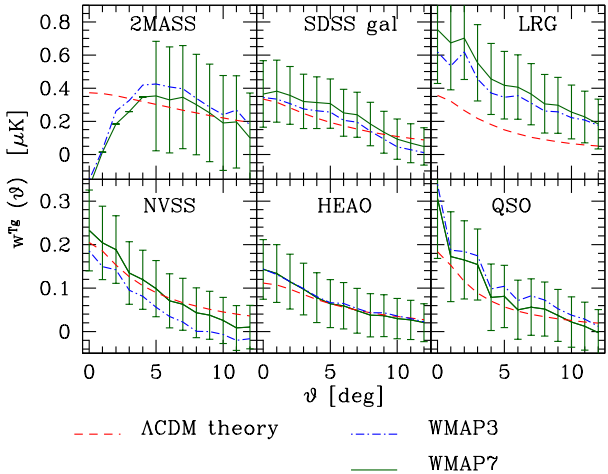}
\caption{Dependence of the results on the different WMAP data releases. Most results vary little compared with the size of the error bars; the largest changes can be seen in NVSS, bringing the results closer to the \LCDM~expectations. The error bars on 2MASS are again 0.5-$\sigma$.} 
\label{fig:W357}
\end{figure}
\end{center}

{The only CCFs for which we see a non-negligible change in Fig.~\ref{fig:wmap7} compared to the earlier analysis by G08 are the LRGs and the NVSS, where the signal has somewhat increased}.  This appears to be primarily due to changes in the WMAP data rather than in the LSS surveys; in Fig.~\ref{fig:W357} we show the CCFs resulting when the current LSS maps are correlated with different WMAP data releases.  
With the exception of the LRGs, the changes tend to bring the data into better agreement with the \LCDM~theory. 
We have found that similar changes appear if the single-frequency, cleaner maps (V and W) are used instead of the ILC. Further, we found that a significant part of these changes is induced by the change in the WMAP mask between the different releases rather than the change in the data themselves, suggesting that the differences may be due to a better foreground cleaning.

To provide a global estimate of the combined significance, we use a theory template $\bar w^{Tg} (\vartheta_i) =  A  g (\vartheta_i)$, where $g (\vartheta_i) \equiv g_i $ is the theoretical prediction of the WMAP7 best fit
model and $A$ is the fit amplitude for each catalogue; further details can be found in G08. By analytically maximising the likelihood, we obtain that the best value  $A$ and its variance for each catalogue are:
\be
A = \frac {\sum_{i,j=1}^p {{\mathcal {C}}_{ij}^{-1} g_i \hat w^{Tg}_j } }{\sum_{i,j=1}^p {\mathcal {C}}_{ij}^{-1} g_i g_j} \, , \hspace{2cm} {\sigma^2_A} = {\left[ \sum_{i,j=1}^p {\mathcal {C}}_{ij}^{-1} g_i g_j  \right]^{-1}} \, ,
\ee
where $\hat w^{Tg}_i$ are the observed CCF for each survey (sampled in $p=13$ angular bins) and ${\mathcal {C}}_{ij}$ is the measured covariance matrix of dimension $p$ described above.
To obtain an unbiased estimator of the inverse covariance ${\mathcal {C}}_{ij}^{-1}$, we correct the result obtained by inverting ${\mathcal {C}}_{ij}$ by a factor $\alpha = {(N-p-2)}/{(N-1)}$ \citep{2007A&A...464..399H}; however in our case (for $N=5000$ realisations) this correction is negligibly small.
This method can be immediately generalised to the full case, in which we fit a single amplitude to a template which includes the six CCFs. In this case the total number of angular bins, and thus the dimension of the covariance matrix, becomes $p = 6 \times 13 = 78$.

\begin {table}
\begin {center}
\begin{tabular}{l | c | c | c}
\hline
Catalogue    &  $A \pm \sigma_A $   &  $S/N$  &  expected $S/N$ \\
\hline
2MASS cut    & 1.40 $\pm$ 2.09   & $ 0.7  $ &  0.5  \\
SDSS gal DR8    & 1.24 $\pm$ 0.57   & $ 2.2  $ & 1.6 \\
SDSS LRG DR7     & 2.10 $\pm$ 0.84   & $ 2.5  $ &  1.2 \\
\hline
NVSS        & 1.21 $\pm$ 0.43  & $ 2.8  $ &  2.6 \\
HEAO        & 1.37 $\pm$ 0.56   & $ 2.4 $  & 2.0  \\
SDSS QSO DR6     & 1.43 $\pm$ 0.62   & $ 2.3  $ & 1.7 \\
\hline
\textbf{TOTAL} & \textbf{1.38 $\pm$ 0.32} & $\mathbf{4.4 \pm 0.4} $ &  $\mathbf{\simeq 3.1}$, $\mathbf{< 7.6}$ \\
\hline
\end{tabular}
\caption{Results from the updated data set compared with the expected signal-to-noise calculated using Eq.~(\ref{eq:snreal}) for each catalogue. The first five data points for 2MASS have been excluded. For the total expected $S/N$, we show both the value estimated from the MC mocks (see below), and using the upper limit of Eq.~(\ref{eq:snmax}). We estimate a $\sim 0.4 \, \sigma$ systematic error on the total signal-to-noise ratio, due to possible different masks and other choices entering into its determination.}
\label{tab:results}
\end{center}
\end {table}

The results with this method and the new data are given in Table~\ref{tab:results}, where we can see that if we identify the signal-to-noise ratio as $S/N = A / \sigma_A$, then the total significance of a detection is now at the $4.4 \, \sigma$ level when a single amplitude is used for all six catalogues.

 It is worth noticing that our significance estimation is however
based on a fiducial model which includes not only the WMAP7 best-fit parameters, but also the assumed redshift distributions and simple bias model of the surveys. While this is reasonable to give an initial estimate of the significance of the detection, a full cosmological analysis should ideally take into account the uncertainties in these quantities, e.g. with the help of additional nuisance parameters. The assumption of constant biases is especially uncertain, in particular for very deep catalogues like HEAO and NVSS \citep[see e.g.][]{2009MNRAS.397..925S}, and this issue should be addressed in a full cosmological analysis allowing for a more realistic bias evolution.
The different assumptions for the biases and for the redshift distributions may for example explain the difference between our results and those by \citet{Ho:2008bz}, where a higher excess signal was found, at the $2 \, \sigma$ level above the \LCDM~predictions.

\begin{figure}
\begin{center}
\includegraphics[width=\linewidth, angle=0]{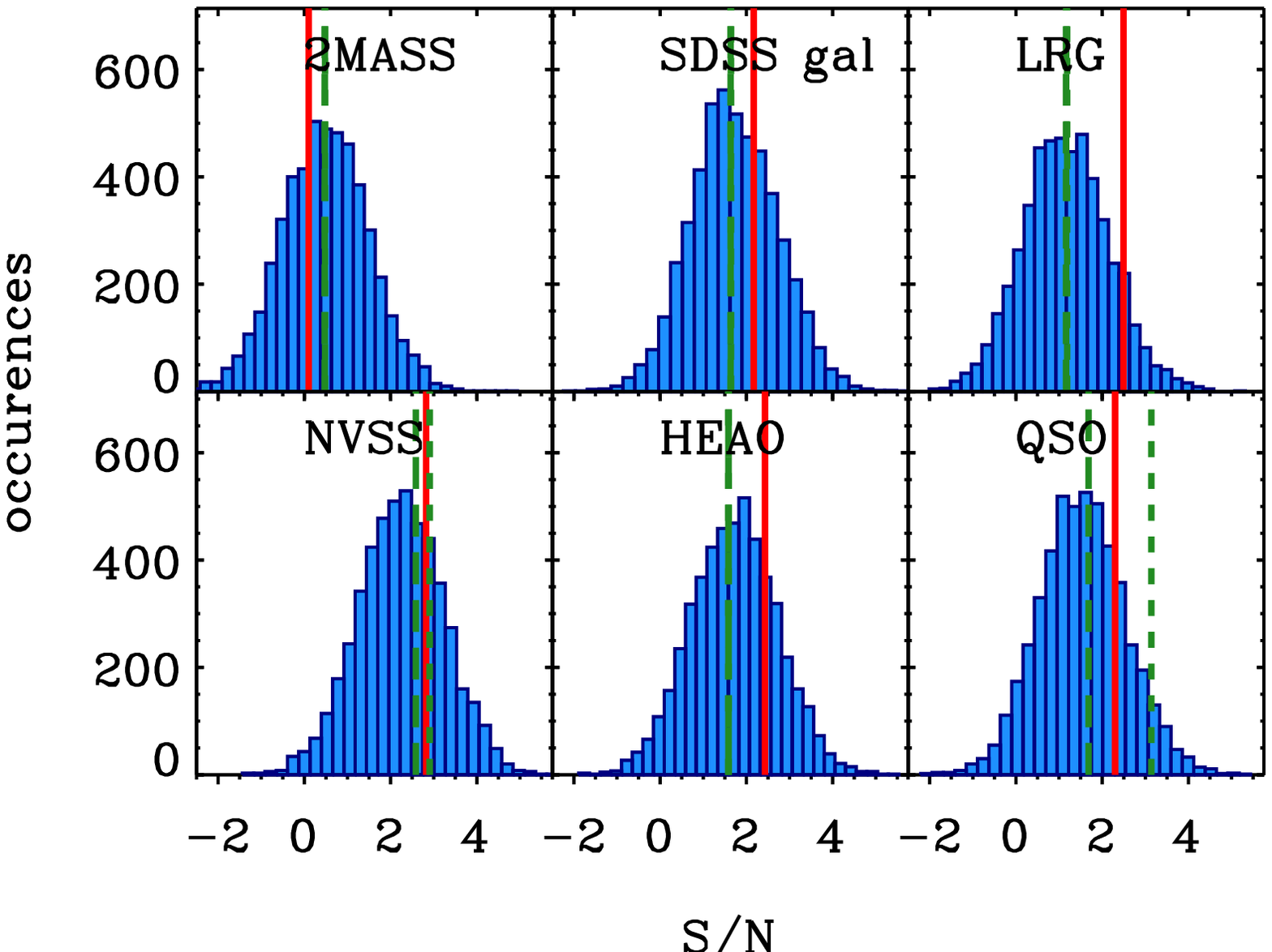}
\includegraphics[width=\linewidth, angle=0]{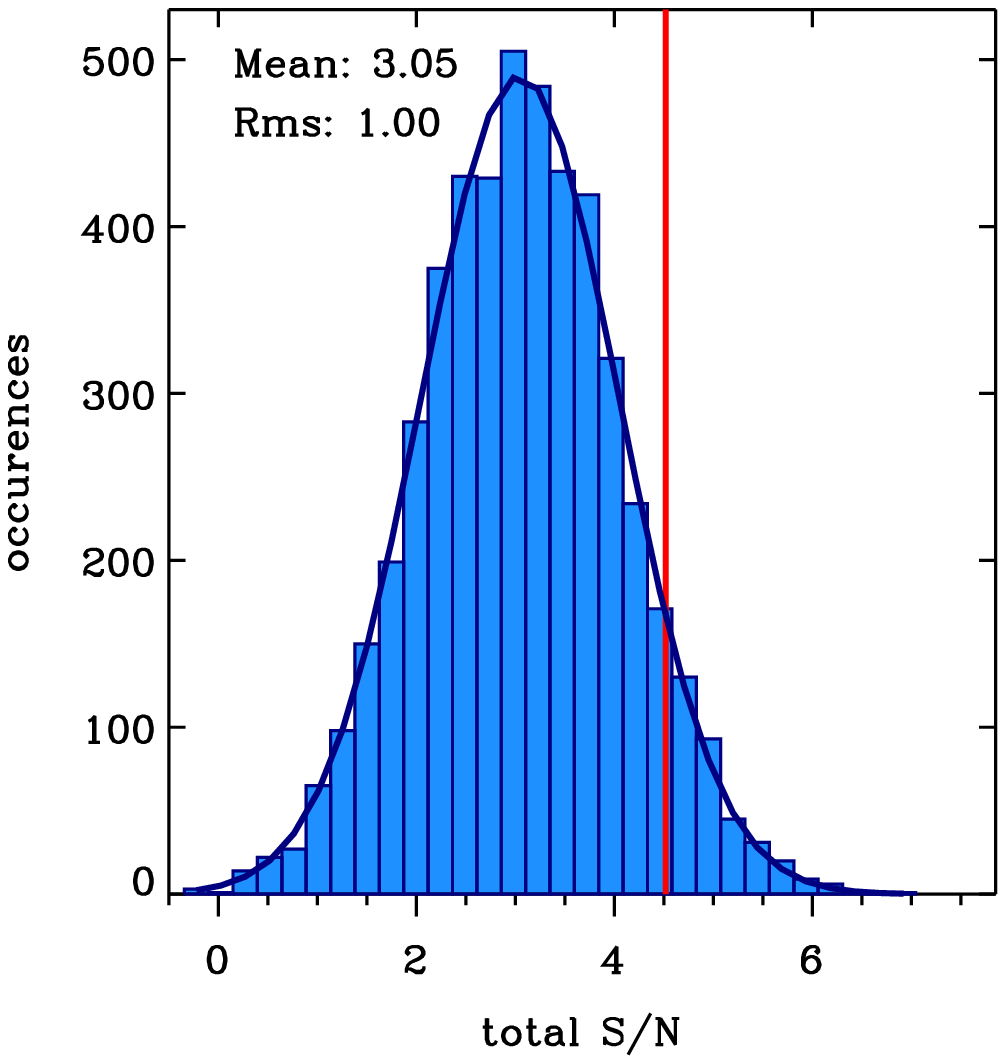}
\caption{Distribution of the signal-to-noise ratio for our 5000 MC realisations (blue histograms), compared with the observations (red solid lines) and the theoretical expectations (green). The different green lines refer to: no shot noise (short-dashed), and shot noise included (long-dashed). The top panel shows each catalogue separately, the bottom panel is the full combination, for which we also show the best-fit Gaussian distribution and its parameters.}
\label{fig:snhisto}
\end{center}
\end{figure}

In Table~\ref{tab:results} we also compare the results with the expected signal-to-noise calculated using Eq.~(\ref{eq:snreal}) for each catalogue, and using the upper limit of Eq.~(\ref{eq:snmax}) for the total. We can see that the measured results are higher than the expectations in most cases.
Given this discrepancy exists between expectations and observations, we next proceed to quantify its significance by studying the distribution of the ISW signal-to-noise obtained using our 5000 mock maps of the galaxy surveys. We show these distributions in Fig.~\ref{fig:snhisto}: here we can see that they are broad, and the position of the observed $S/N$ is well within the expected scatter. In more detail, we fit a Gaussian to the distribution of the mock total signal-to-noise, where we find that the mean (i.e. the expectation for the total $S/N$ from \LCDM) is $3.05$, and the r.m.s. is 1 by construction.  This places our observed result $1.35 \, \sigma$ away from the mean.

A further interesting point to be learnt from Fig.~\ref{fig:snhisto} is the comparison between theoretical $S/N$ with (green solid lines) and without (green dashed lines) shot noise. We can see that the effect of shot noise is particularly large for the quasars, due to their limited number density.
 From this we conclude that future measurements of extended quasar catalogues have the potential to significantly improve the existing results, due to the large redshift coverage of these sources.

Given the number of different assumptions in the method of the analysis, we roughly estimate that a systematic uncertainty of $\simeq 0.4$ needs to be included on the final figure of the signal-to-noise ratio.
For example, using other reasonable redshift distributions for the catalogues typically results in changes of the signal-to-noise ratio of the order $0.2 - 0.4$. Similar differences are obtained when changing the thresholds in the extinction masks, or excluding parts of the data (such as the Southern hemisphere for the new SDSS DR8 galaxies). Another change which is typically at the same level is produced if we decide to completely discard the pixels near the edge of the survey, for which the mask weighting is $f^g_i < 1$, instead of correcting them with the appropriate weights.
Furthermore any extra large-scale power in the auto-correlation functions, which could arise e.g. due to low-redshift contamination or other systematics, would increase the variance of the cross-correlations. We discuss this below in Section~\ref{sec:spike}, showing that its effect is limited and in agreement with our systematic estimation of 0.4 $\sigma$.

We have also checked the effect of removing any one catalogue from the analysis, finding in this case that the total significance can not be lowered below $3.9 \, \sigma$, which is the result obtained when ignoring the NVSS data.

We publicly release the maps, masks, and results discussed herein, which can be downloaded from the internet.\footnote{\texttt{www.usm.lmu.de/\~{}tommaso/isw2012.php}}  
Here one can find for each of the six catalogues the following data: a file with the Healpix galaxy map in FITS format, and its companion mask in the same format; a table with the redshift distribution, and a table with the results of the CCFs. Finally, the full covariance matrix based on Monte Carlo maps is also provided.

\section{Systematic uncertainties} \label{sec:systematics}

Since the ISW signal is expected to be weak, possible systematic uncertainties are a serious issue. 
Here we discuss some of the new tests we have explored to constrain this contamination, and further discussion can be found in G08 and 
earlier work. 

\subsection{Foreground contamination}

While individual instrumental systematics are not expected to be correlated between surveys, the cross-correlation could be contaminated by extra-galactic sources in the microwave frequencies, such as synchrotron emission or the Sunyaev-Zel'dovich effect. 
Our galaxy also emits in the microwave (dust, synchrotron and free-free), so it is important to ensure that galactic structure 
does not creep into the large-scale structure measured in surveys; otherwise spurious correlations with the CMB foregrounds will bias 
the cosmological interpretation of the measurements. {Another possible source of cross-correlation is the secondary Doppler effect which may be added to the CMB at the time or reionisation \citep{Giannantonio:2007za}; however, at the redshifts of currently available data, this is expected to be completely subdominant.}

\begin{center}
\begin{figure}
\includegraphics[width=\linewidth, angle=0]{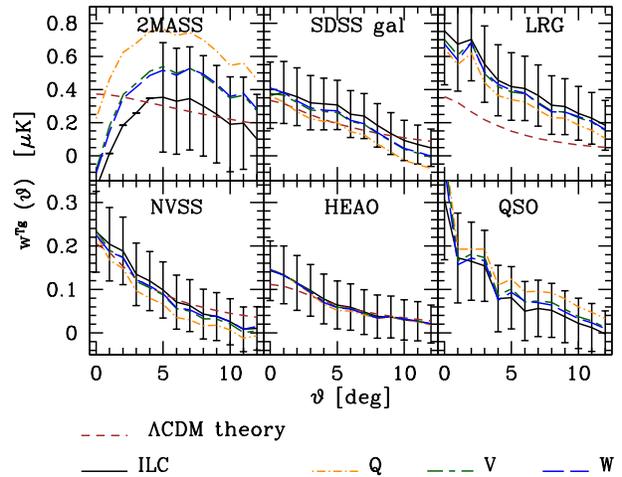}
\caption{Frequency dependence of the cross-correlations, for WMAP7: most results are stable compared with the size of the error bars. The error bars on 2MASS are again 0.5-$\sigma$.} 
\label{fig:freq}
\end{figure}
\end{center}

An important way to keep foreground contamination under control is to check for dependence of the cross-correlation functions on the CMB frequency, as we expect the ISW signal to be monochromatic while the extra-galactic and galactic foregrounds sources such as SZ, synchrotron and dust are expected to have a strong frequency dependence. 
We have performed this test for the new data, and the results can be seen in Fig.~\ref{fig:freq}; the cross-correlation functions 
are quite stable relative to the measurement error bars for the less contaminated WMAP maps (ILC, W, V, and largely Q). The most dependent on frequency is 2MASS, where \citet{Afshordi:2003xu} suggested there was evidence for the SZ effect; the most stable is HEAO, which is perhaps understandable as the hard X-ray background should be least affected by galactic contamination.

Perhaps the most worrying systematic problem which can affect the ISW measurement is the leakage of information from the structure of our Milky Way into the galaxy catalogues. This can potentially jeopardise the results, as it correlates with the CMB via residual dust extinction corrections. This problem can be minimised by masking sky areas closest to the galactic plane, and those areas most affected by reddening, which we have done for all catalogues.

We have also checked for the effect of removing from the maps a band centred on {either the ecliptic or galactic planes}, using cuts of different widths (10, 20 and 30 degrees). We have found no significant differences.
 Other
possibile systematics include the effects of  point
sources, regions of poor seeing and
 sky brightness; as discussed in G08 they are less severe than dust extinction for our data.

For the DR7 MegaZ LRG data set there is excess power in the correlation with the CMB, at a similar level as found with DR6; excess large-scale power has also been seen in the MegaZ LRG auto-correlation function as discussed by \citet{2011PhRvL.106x1301T}.
While these excesses may be cosmological, they could also be evidence of remaining systematics, such as a higher than expected stellar contamination. Recently, \citet{2011MNRAS.417.1350R} improved the methods for selecting and interpreting an LRG catalogue, using a large spectroscopic training set from SDSS-III BOSS \citep{Eis11}. These authors  found that the most serious systematic problem with present photometric redshift catalogues from SDSS imaging is the failure to detect faint galaxies around foreground stars (even relatively faint stars to $r\simeq20$).  \citet{2011MNRAS.417.1350R} corrected this issue, but it does illustrate the need to be diligent about stellar contamination, especially when cross-correlating with other datasets which may also include some contamination with galactic sources. 
In the future, catalogues with this higher level of systematics control should be used for the measurement of the ISW. 
The ISW signal seen in the LRG cross-correlation is higher than the \LCDM~expectation; if future LRG data will be more in accord with \LCDM, the best-fit amplitude may well decline, but with the uncertainty also shrinking, the signal-to-noise ratio could remain at a similar level.  

\subsection {The rotation test} \label{sec:rotation}

Although the presence of systematic effects has been
studied rather carefully by several authors, it is  possible  some unaccounted
 uncertainty could remain in the data, thus compromising the measurements,
and it is therefore worthwhile to look for new ways to check
for such problems. 

One such test, based on arbitrary rotations of the sky maps, was recently proposed by \cite{2010MNRAS.402.2228S}.
In this test, cross-correlation functions (CCF) are generated by cross-correlating the true maps, but after one of the maps has been 
\textit{rotated by some arbitrary angle} $\Delta \varphi$ relative to the other. In
\cite{2010MNRAS.402.2228S} rotations around the Galactic axis were
chosen particularly to try to identify galactic contamination, but any axis choice is possible. For a sufficiently large rotation $\Delta \varphi$, one expects on average that there will be no correlation, though any particular measurement will have scatter reflecting the intrinsic variance in the measurement.  (Any rotation leaves two poles fixed, implying a small amount of residual correlation, but in practice this is negligible.)

The critical question is, how do we evaluate the significance of the rotated correlation functions?   The most obvious comparison is to the variance of the intrinsic scatter, inferred from our Monte Carlo simulations.  Are the rotated correlation measurements consistent with this intrinsic scatter or not? 

\begin{center}
\begin{figure}
\includegraphics[width=\linewidth, angle=0]{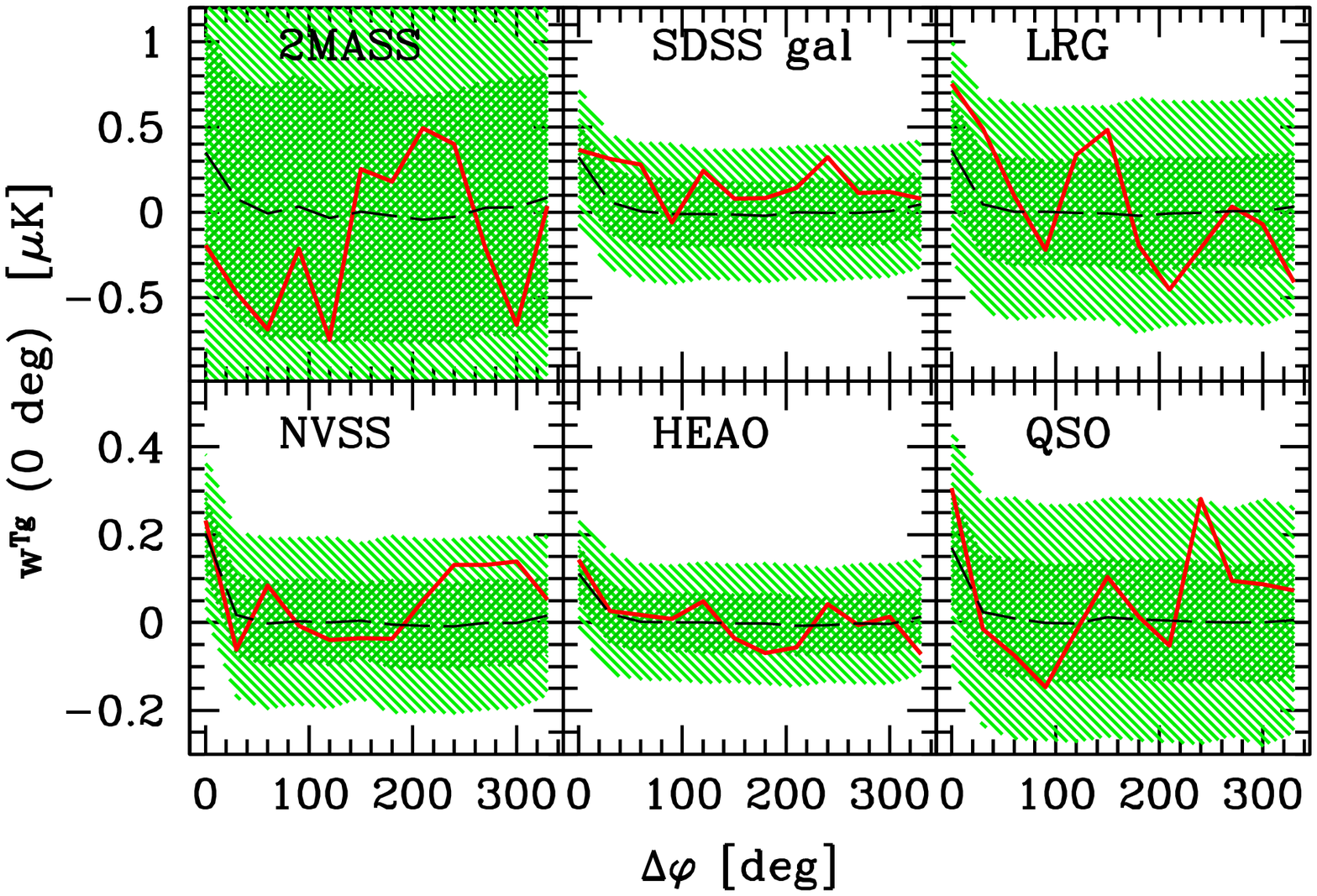}
\includegraphics[width=\linewidth, angle=0]{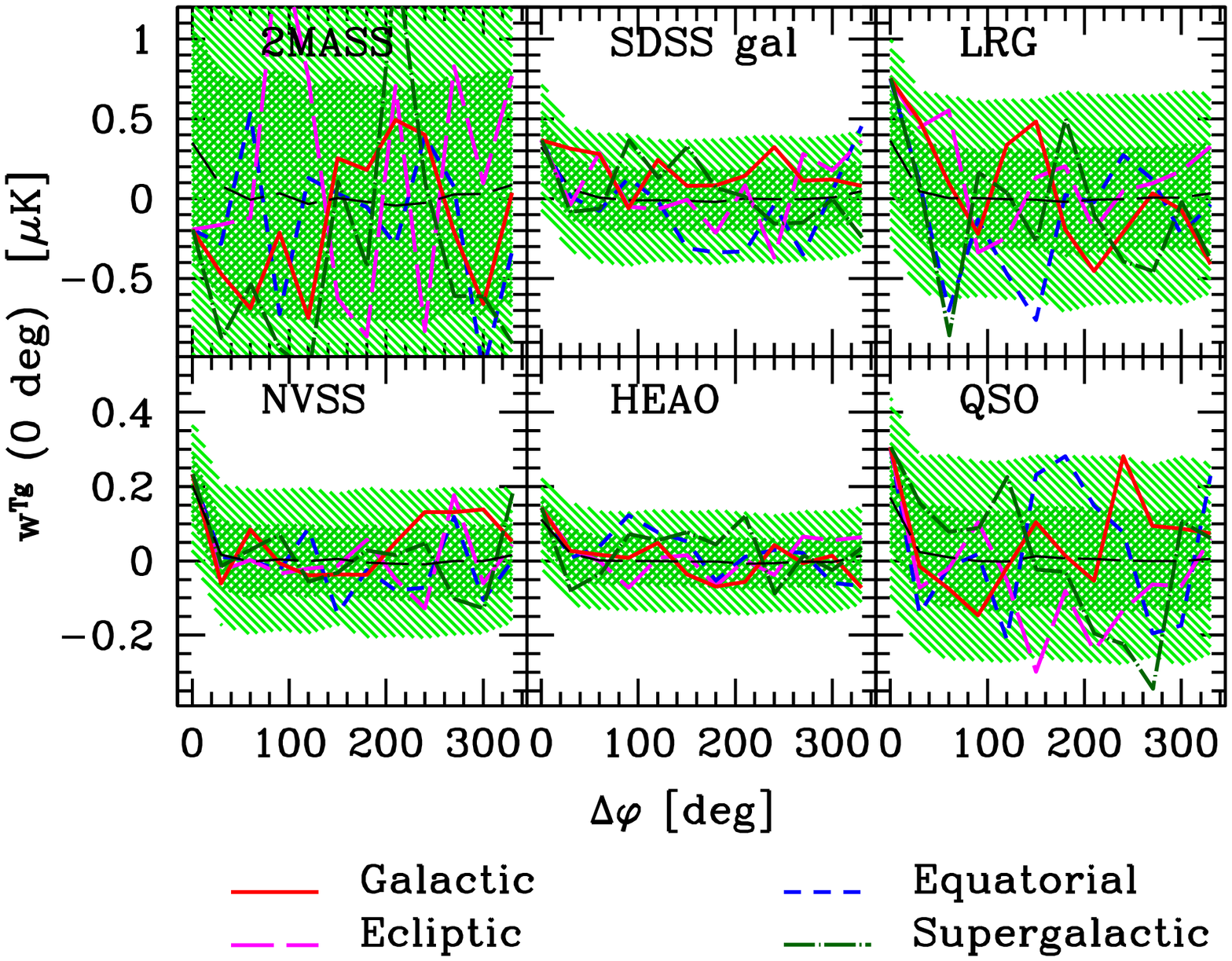}
\caption{The rotation test for our data, zero-lag case. In the top panel, for each galaxy catalogue, we show the
  evolution of the CCF at 0 degrees as a function of the arbitrary rotation
  angle $\Delta \varphi$, describing rotations around the galactic
  plane. The coloured lines show the observed results; the green shaded bands
  indicate 1 and $2 \, \sigma$ regions calculated by generating 500 mock
  Monte Carlo maps of the data. The black dashed lines are the averages seen 
  in the same mock data, assuming a \LCDM~concordance model. The bottom panel shows the same test for rotations in  different coordinate systems: galactic, equatorial, ecliptic and supergalactic.}
\label{fig:rotshadeC0}
\end{figure}
\end{center}

\subsubsection{Assessing the significance}

One difficulty in making this comparison is that, for a given rotation axis, there are a limited number of rotations which are independent of one another due to the fact that the ISW signal is on relatively large angles.   We use 500 Monte Carlo simulations to evaluate this; we create sets of CMB and galaxy maps with the expected signal for the \LCDM~model, and rotate them as we do the real data.  In the top panel of Fig.~\ref{fig:rotshadeC0} we show the  average `zero-lag' cross-correlation as a function of rotation angle, with the 1-$\sigma$ and 2-$\sigma$ regions shaded in green.  We see that this is significant typically out to 30 degrees, implying that this is the minimum rotation required to provide an independent sample. 

For any given rotation axis, this leaves only 11 independent samples of the cross-correlation measure.  
If one or more of these are significantly higher the the r.m.s. amplitude, then this could be evidence for the systematic contamination.   
In the top panel of Fig.~\ref{fig:rotshadeC0} we plot the `zero-lag' cross-correlation for each of our catalogues.  
This shows that no significant outliers exist, and that the signal is generally highest when there is no rotation, apart from 2MASS where the 
zero-lag signal could be cancelled by the SZ effect.  
This is quantified in Table~\ref{tab:oursc0}, where we count the number of rotations exceeding thresholds of 1-, 2$\sigma$ (expected 32\% and 5\%); if anything, the number of high correlations seen in the rotated maps is lower than expected, though this discrepency is not significant. The observed number above the threshold should be binomially distributed, which is very broad given the limited number of observations.

In S10, they looked instead at the number of rotations where the  `zero-lag' cross-correlation exceeded the true `zero-lag' cross-correlation for that particular survey; that is, our zero degree value (denoted $w_{0i}$).  However, this is just one choice, with the drawback of being different for each data set $i$. Nonetheless, we also show this in Table~\ref{tab:oursc0}.  Again, for rotations about the galactic axis, we find none exceeding the true value for our data sets.  

\begin{table*}
\begin{tabular}{l | c c c c c c} 
\hline
                & \multicolumn{6}{c}{Zero-lag data only}   \\ 
\hline
Catalogue & $>1 \sigma$ $ \ne 0 $ & -- multi-axes  & $>2 \sigma$ $ \ne 0 $ & -- multi-axes &  $> w_{0i} $  & -- multi-axes  \\ 
\hline
2MASS     &      0/11  ($ 0\%$)   & 12/44 ($27\%$) &    0/11  ($ 0\%$)     & 2/44   ($5\%$)&     ---      &       ---      \\   
SDSS gal  &      4/11  ($36\%$)   & 16/44 ($36\%$) &    0/11  ($ 0\%$)     & 1/44   ($2\%$)& 0/11 ($0\%$) &  3/44 ($7\%$)  \\ 
SDSS LRG  &      5/11  ($45\%$)   & 16/44 ($36\%$) &    0/11  ($ 0\%$)     & 3/44   ($7\%$)& 0/11 ($0\%$) &  2/44 ($5\%$)  \\ 
\hline
NVSS      &      3/11  ($27\%$)   & 11/44 ($25\%$) &    0/11  ($ 0\%$)     & 0/44   ($0\%$)& 0/11 ($0\%$) &  0/44 ($0\%$)  \\ 
HEAO      &      1/11  ($ 9\%$)   & 10/44 ($23\%$) &    0/11  ($ 0\%$)     & 0/44   ($0\%$)& 0/11 ($0\%$) &  0/44 ($0\%$)  \\ 
SDSS QSO  &      2/11  ($18\%$)   & 17/44 ($39\%$) &    1/11  ($ 9\%$)     & 3/44   ($7\%$)& 0/11 ($0\%$) &  1/44 ($2\%$)  \\ 
\hline
Total     &     15/66  ($23\%$)   & 82/264 ($31\%$)&    1/66  ($ 2\%$)     & 9/264  ($3\%$)& 0/55 ($0\%$) &  6/220 ($3\%$) \\ 
Expected  &     21/66  ($32\%$)   & 84/264 ($32\%$)&    3/66  ($ 5\%$)     & 13/264 ($5\%$)&       ---    &  ---           \\ 
\hline
\end{tabular}
\caption{Results of the rotation test with our data for the zero-lag CCF. The scatter is
   in reasonably good agreement with the expectations, and is smaller than in S10 data (see Table~\ref{tab:their1}). In the rotations around the galactic axis there are fewer outliers than expected, but the numbers become closer to the expectations when using a larger number of rotation axes.}
\label{tab:oursc0}
\end{table*}

\begin{table*}
\begin{tabular}{l  | c c c c c c|}
\hline
                  & \multicolumn{6}{c}{Template amplitude fitting}  \\
\hline                                                                                                                 
Catalogue &     $>1 \sigma$ $ \ne 0 $ &  -- multi-axes    & $>2 \sigma$ $ \ne 0 $ &  -- multi-axes &  $> A_i $  & -- multi-axes \\
\hline
2MASS     &     0/11  ($  0\%$)       & 12/44  ($ 27\%$)  &  0/11  ($ 0\%$)       &  3/44  ($ 7\%$) &    ---      &     ---       \\
SDSS gal  &     4/11  ($ 36\%$)       & 17/44  ($ 39\%$)  &  0/11  ($ 0\%$)       &  3/44  ($ 7\%$) & 0/11 ($0\%$)&  2/44 ($5\%$) \\
SDSS LRG  &     6/11  ($ 55\%$)       & 17/44  ($ 39\%$)  &  1/11  ($ 9\%$)       &  2/44  ($ 5\%$) & 0/11 ($0\%$)&  0/44 ($0\%$) \\
\hline
NVSS      &     3/11  ($ 27\%$)       & 10/44  ($ 23\%$)  &  0/11  ($ 0\%$)       &  2/44  ($ 5\%$) & 0/11 ($0\%$)&  0/44 ($0\%$) \\
HEAO      &     4/11  ($ 36\%$)       & 15/44  ($ 34\%$)  &  0/11  ($ 0\%$)       &  0/44  ($ 0\%$) & 0/11 ($0\%$)&  0/44 ($0\%$) \\
SDSS QSO  &     3/11  ($ 27\%$)       & 18/44  ($ 41\%$)  &  1/11  ($ 9\%$)       &  3/44  ($ 7\%$) & 1/11 ($9\%$)&  3/44 ($7\%$) \\
\hline
Total     &    20/66  ($ 30\%$)       & 89/264 ($ 34\%$)  &  2/66  ($ 3\%$)       & 13/264 ($ 5\%$) & 1/55 ($2\%$)& 5/220 ($2\%$) \\
Expected  &    21/66  ($ 32\%$)       & 84/264 ($ 32\%$)  &  3/66  ($ 5\%$)       & 13/264 ($ 5\%$) &     ---     &      ---      \\ 
\hline
\end{tabular}
\caption{As above, for the template amplitude fitting. The scatter is here
   in even better agreement with the expectations.}
\label{tab:oursA}
\end{table*}

We can increase our statistics and explore for the possibility of systematics associated with other reference frames by considering rotations about other axes.  We have  repeated the test with rotations using:
 galactic, ecliptical, equatorial and supergalactic axes.  For the discussion below we ignore chance alignments of the various rotations, but it should be kept in mind that all these points may not be fully independent.
The results are shown in the bottom panel of  Fig.~\ref{fig:rotshadeC0}.
\textit{All of the rotations are consistent with the expected scatter, showing no indication for a preferred rotation axis.}  This seems to support the hypothesis that the rotations are merely providing a measure of the intrinsic scatter and are not suggestive of a contaminant associated with a single axis.  The statistics are summarised in Table~\ref{tab:oursc0}, where we can see that only 31\% and 3\%  of the points lie above the 1-, 2-$\sigma$ thresholds respectively. Excluding 2MASS, only 3\% exceed the true value measured in the data; unfortunately, this statistic cannot be used directly to estimate a total significance because it will be dominated by maps with the least significant amplitude.  

\begin{center}
\begin{figure}
\includegraphics[width=0.99\linewidth, angle=0]{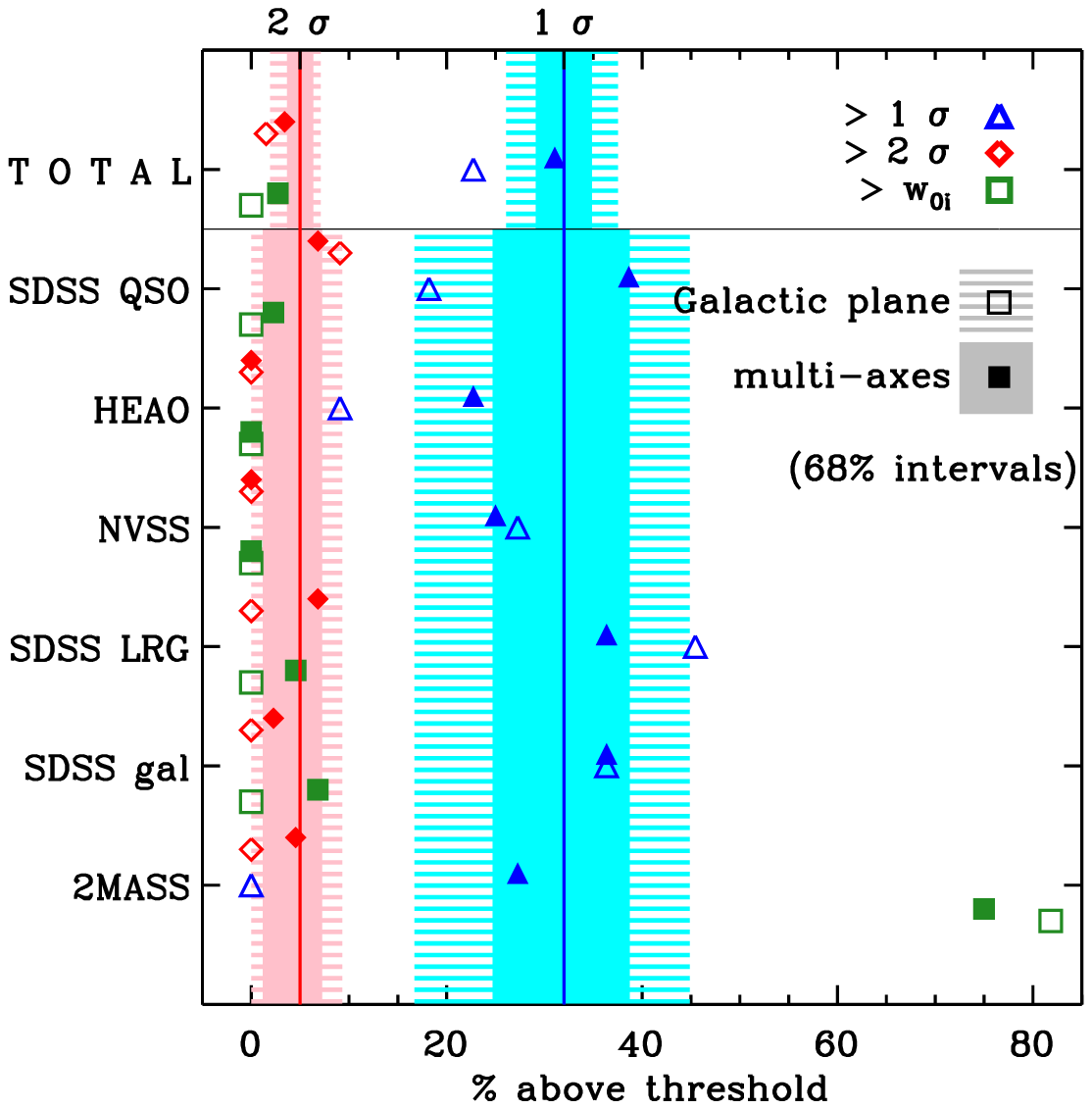} 
\includegraphics[width=0.99\linewidth, angle=0]{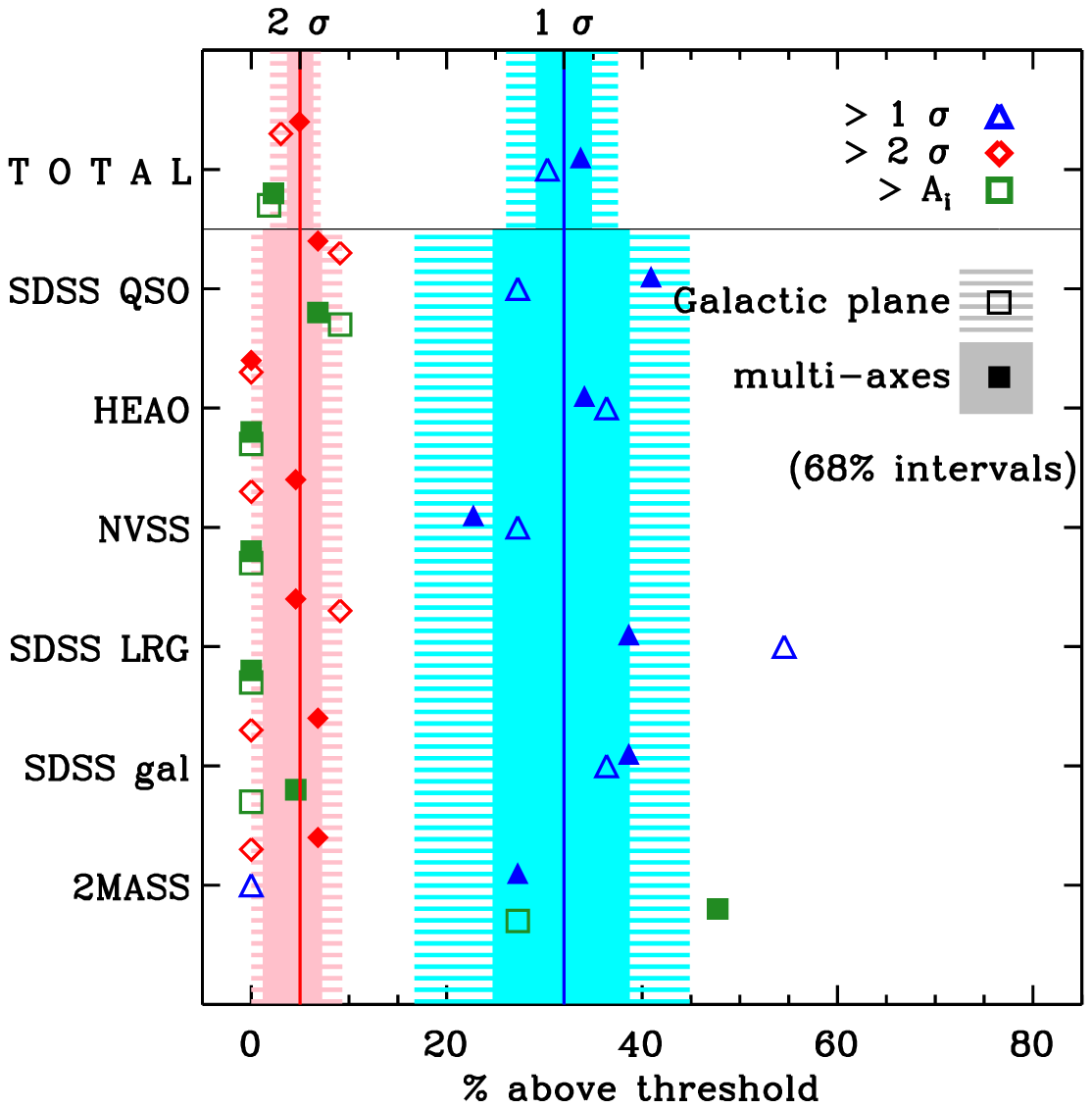}  
\caption{Significance of the rotation test for our combined data, using WMAP7, zero-lag only (above) and full template amplitude fitting (below). The data points correspond to the percentage of realisations observed above any given threshold: $1 \, \sigma$ (blue triangles), $2 \, \sigma$ (red diamonds), and each catalogue's unrotated CCF $w_{0i}$ or amplitude $A_{i}$ (green squares). These should be compared to the  expected 1- and 2-$\sigma$ fraction (vertical lines); the error bands are $68\%$ confidence intervals assuming a binomial probability distribution for the counts. The empty points and the dashed bands refer to rotations around the Galactic plane only, while the filled points and solid bands include data from all four rotation axes.}
\label{fig:PoissG08}
\end{figure}
\end{center}

For a visual impression of this, we plot our results in Fig.~\ref{fig:PoissG08}, where the observed number of detections in excess of each threshold is compared with the $68\%$ confidence regions drawn from the continuous generalisation of a binomial probability distribution.
 Here we can appreciate that the distribution of the points above the 1- and 2-$\sigma$ thresholds is fully consistent with the expected scatter and the number of points above the the true measure levels  seems consistent with an average level of significance $\sim 2 \, \sigma$ for each catalogue.

Thus we find no evidence for systematic contamination from the rotation tests and that the rotated cross-correlation measurements are consistent with those expected from the covariance matrix derived from Monte Carlo simulations.  Indeed, the rotations are an alternative means of deriving the covariance and were used in early studies as a means of calculating it.  Unfortunately, the number of possible independent rotations is limited, meaning it is difficult to derive a stable covariance from rotations alone. 

\subsubsection{Including full angular information}

\begin{center}
\begin{figure}
\includegraphics[width=\linewidth, angle=0]{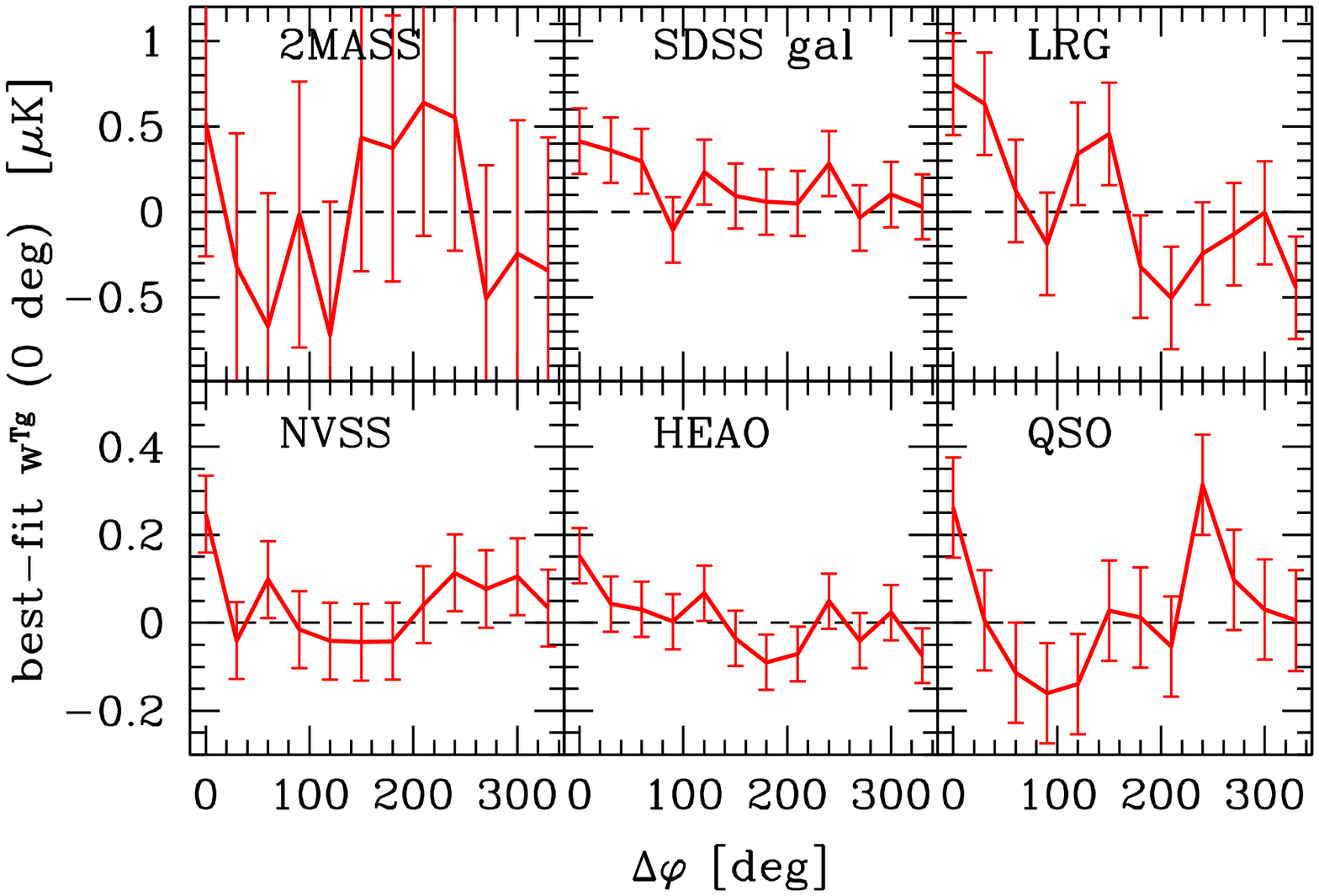}
\includegraphics[width=\linewidth, angle=0]{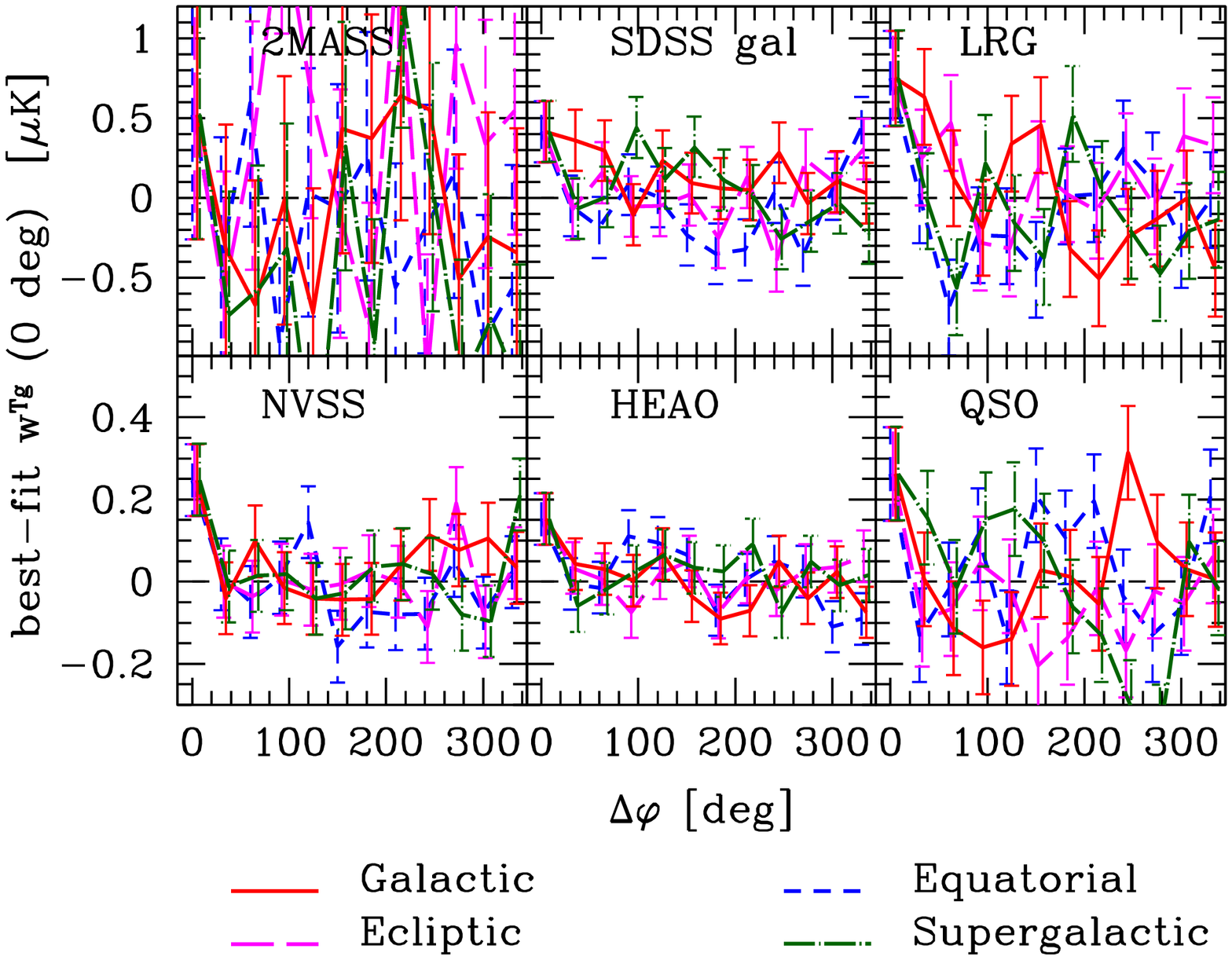}
\caption{The rotation test for our data, best amplitude case.  In this case we show in the top panel the best fit amplitude, using all the angular information, and assuming a templated based on the best fit WMAP7 \LCDM~cosmology; the bottom panel shows the same using multiple rotation axes: Galactic, Equatorial, Ecliptic and Supergalactic.} 
\label{fig:rotshadeA}
\end{figure}
\end{center}

For the discussion above, we have focused on the zero-lag cross-correlation following S10.  However, this single bin contains only part of the ISW signal, and thus the significance derived from it is not as high as one gets when including the full cross-correlation function.  For this reason, we repeat our analysis fitting to templates based on the predicted correlations expected in the WMAP7 best-fit cosmology.  
We use the covariance matrices obtained with the expected CCF; we plot the best-fit amplitudes of the rotated maps as a function of the rotation angle in the first panel of Fig.~\ref{fig:rotshadeA}. 
The scatter of the $A$'s from the rotated maps is in even better agreement with expectations, as can be seen in Table~\ref{tab:oursA}. Here, we see that only 30\% and 3\% of the points lie above the thresholds of $1 \, \sigma$ and $2 \, \sigma$.  2\% exceed the the level of the unrotated best-fit amplitude, denoted as $A_{i}$.  

We have also performed the multi-axes test in the amplitude fitting case, as shown in the second panel of Fig.~\ref{fig:rotshadeA}. We can see in the figure, and in the results summarised in Table~\ref{tab:oursA}, that the results remain consistent over an increased population. 

The significance of these results can be  better appreciated by considering the uncertainties on the number counts, shown in the lower panel of Fig.~\ref{fig:PoissG08}. Here we can see once again that not only are the 1- and 2-$\sigma$ thresholds consistent with the distribution of the outlying data, but also the populations above the $A_{i}$ thresholds are consistent with a level of significance $> 2 \, \sigma$ for each catalogue (excluding the 2MASS data, where the detection has low significance).

\subsubsection {Application to \citet{2010MNRAS.402.2228S} data} \label{sec:S09rots}
We have shown above that for our data, the results of the rotation test are in agreement with the Monte Carlo estimations of the variance; let us now discuss the application to the S10 data themselves.  It was claimed by \cite{2010MNRAS.402.2228S} that the result of this test for multiple data sets (SDSS gal, NVSS, SDSS LRGs and
AAOmega LRGs) was in contradiction with the claimed detection of the ISW, pointing instead towards strong unknown residual systematics.

By looking at Figure 14 in S10, we compile the statistics shown in Table~\ref{tab:their1}.  Here we can see that in total $39\%$ of the random points obtained by arbitrary rotations are scattered at $> 1 \, \sigma$ away from zero, while randomly we would expect this fraction to be $32\%$.
If we instead choose the 2-$\sigma$ threshold, we find a total of $6/56$ points above this threshold, corresponding to $11\%$ compared to the expectation of only $5\%$.
Alternatively following S10, we can choose the threshold to be the unrotated signal $w_{0i}$ seen in the data; in this case we find that 13/48 points (27\%) are above this mark, if we discard the points from AAOmega for which S10 find no correlation.

\begin{center}
\begin{figure}
\includegraphics[width=0.99\linewidth, angle=0]{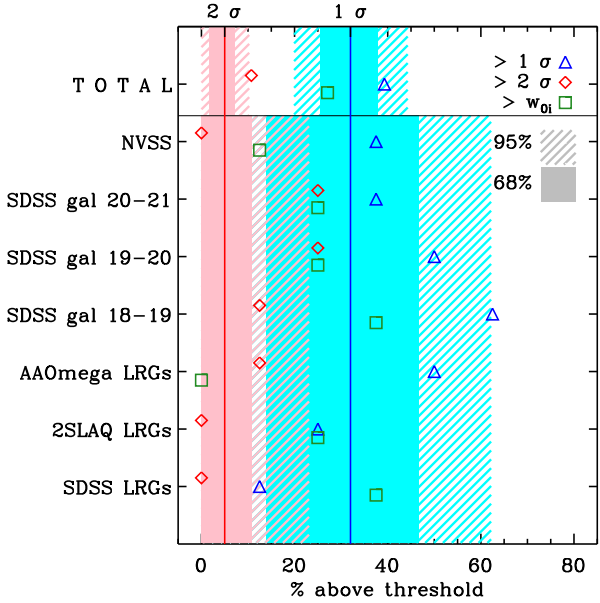} 
\caption{Significance of the rotation test for the data of S10, for each individual catalogue and (on top) for the total. The data points correspond to the percentage of realisations observed above any given threshold: $1 \, \sigma$ (blue triangles), $2 \, \sigma$ (red diamonds), and each catalogue's unrotated CCF $w_{0i}$ (green squares). The expected 1- and 2-$\sigma$ thresholds are overplotted with vertical lines. {The error bands in this case are $68\%$ (solid) and $95\%$ (dashed) confidence intervals assuming a binomial probability distribution. }}
\label{fig:PossS09}
\end{figure}
\end{center}

By including the expected errors on the counts of the outlying points, we can see the uncertainties corresponding to the data by S10 in Fig.~\ref{fig:PossS09}. Here we plot, for each data set and for the total, the number of bins above each threshold of $1 \, \sigma$, $2 \, \sigma$, and the `real' value of the CCF at no rotation $w_{0i}$ (again, excluding the AAOmega data for this last case since here $w_{0i}$ is not defined). 
We can see here a higher scatter than in our data, but this not unexpected given the lower number of realisation. By comparing the data with the $68\%$ and $95\%$ confidence intervals from the binomial distribution, we observe substantial agreement.

With respect to the $w_{0i}$ threshold, Fig.~\ref{fig:PossS09} indicates that NVSS and, to a lesser extent the SDSS 20-21, 19-20, and 2SLAQ LRGs are consistent with a significance level $>2 \, \sigma$, while the other catalogues seem to have a lower significance level. 
Also, it should be mentioned that the error bars were calculated using a jack-knife method, which can often depend on the details of how the jack-knife was performed.  In earlier studies, jack-knife error estimates have been seen to be somewhat smaller than those inferred from 
 Monte Carlo methods \citep{2007MNRAS.381.1347C}. 
 This could explain why somewhat more of the random rotations appear significant in the S10 data compared to our analysis.  
 Overall however, when considered with the results based on our maps, we find that \textit{the rotation test provides no significant evidence for residual systematics in the data}.

\begin{table}
\begin{tabular}{ l c c c c c c}
\hline
Catalogue           &  \multicolumn{2}{c}{ $>1 \, \sigma$  $ \ne 0$}    &   \multicolumn{2}{c}{ $>2 \, \sigma$  $ \ne 0$ } &   \multicolumn{2}{c}{ $>w_{0i}$  }   \\
\hline
SDSS LRGs                 &  1/8  &   $ 13 \% $ &  0/8  &   $  0 \% $ & 3/8 & $38\%$ \\
2SLAQ  LRGs               &  2/8  &   $ 25 \% $ &  0/8  &   $  0 \% $ & 2/8 & $25\%$ \\
AAOmega LRGs              &  4/8  &   $ 50 \% $ &  1/8  &   $ 13 \% $ & --- & --- \\ 
SDSS gal $18<r<19$        &  5/8  &   $ 63 \% $ &  1/8  &   $ 13 \% $ & 3/8 & $38\%$ \\ 
SDSS gal $19<r<20$        &  4/8  &   $ 50 \% $ &  2/8  &   $ 25 \% $ & 2/8 & $25\%$ \\ 
SDSS gal $20<r<21$        &  3/8  &   $ 38 \% $ &  2/8  &   $ 25 \% $ & 2/8 & $25\%$ \\ 
NVSS                      &  3/8  &   $ 38 \% $ &  0/8  &   $  0 \% $ & 1/8 & $13\%$ \\ 
\hline
Total                     & 22/56 &   $ 39 \% $ &  6/56  &   $ 11 \% $ & 13/48 & $27\%$ \\
Expected                  & 18/56 &   $ 32 \% $ &  3/56  &   $ 5 \% $ &  --- & --- \\ 
\hline
\end{tabular}
\caption{Results of the rotation test in \citet{2010MNRAS.402.2228S}. By definition, we would expect $32 \%$ ($5 \%$) of the  points to be further than $1 \, \sigma$ ($2 \, \sigma$) away from zero.}
\label{tab:their1}
\end{table}

\section{Discussion} \label{sec:discussion}

A handful of other papers have appeared with results that seem to be in conflict with the more common approaches described above.  The differences are centred on three main issues: large differences in the statistical significance of the results (with papers suggesting both higher and lower significance),  questions related to new data sets and questions about systematic contaminations.  
We discuss each of these in turn.  

\subsection {Statistical significance}

\subsubsection{Theoretical covariance}  \label{sec:comparecvm}
The significance of the measurements is estimated given the full covariance between the data. This may be obtained using different methods: analytically (``theory'' covariance), from the data themselves (jack-knife or bootstrapping), or using Monte Carlos, which is our method of choice. It was shown already by \citet{2007MNRAS.381.1347C} that all such methods should give consistent results (with fluctuations of order $10 \%$ between them), and G08 presented a comparison between MC and jack-knife (JK) covariances showing that the results were comparable, although the JK method was deemed less stable.
Here we compare our baseline MC covariance with the theory (TH) covariance calculated analytically. Extending the derivations by \citet{2007MNRAS.381.1347C,2009MNRAS.397.1348W,2011MNRAS.415.2193R}, we find for the element of the covariance between two galaxy catalogues $i,j$ in two angular bins $\vartheta,\vartheta'$:
\ba
&~& \mathcal{C}^{\mathrm{TH}}_{ij} (\vartheta,\vartheta') \equiv  \mathrm{Cov} \left[ w^{Tg_i} (\vartheta) \, , w^{Tg_j} (\vartheta') \right]    =  \\
 &=&  \sum_l \frac{\left(2 l + 1 \right)^2}  {\left( 4  \pi \right)^2   \sqrt{f_{\mathrm{sky}}^i \, f_{\mathrm{sky}}^j  }}  \,  P_l\left[ \cos (\vartheta) \right] \,  P_l\left[ \cos (\vartheta') \right]  \, \mathrm{Cov} \left(C_l^{Tg_i} , C_l^{Tg_j}  \right) \, , \nonumber 
\ea
where $P_l$ are the Legendre polynomials and the harmonic space covariance is given by
\be
\mathrm{Cov} \left(C_l^{Tg_i} , C_l^{Tg_j}  \right) = \frac{1}{2 l + 1} \left[ C_l^{Tg_i} \, C_l^{Tg_j} + C_l^{TT, \, \mathrm{tot}}  \left( C_l^{g_i g_j} + \delta_{ij} / n_{s,i}  \right)   \right] \, ,
\ee
where $\delta_{ij}$ is the Kronecker symbol.

We find that the analytical covariance is comparable with the MC method used in our main analysis; the diagonal elements agree at the $\sim 10 \%$ level. The significance of the detections is decreased in this case by $\sim 10 \%$:
 we find for the total combined result $A_{\mathrm{tot}}^{\mathrm{TH}} = 1.42 \pm 0.34$, i.e. $S/N = 4.1$ when using the theory covariance.
  This is well within the expected fluctuations given the differences in the two procedures.

\subsubsection {Absolute versus relative probability }

When comparing results from different papers, one must be careful to ensure that they are asking the same questions.  As discussed in G08 and S10, there are at least two ways of quantifying the significance of the ISW detection; the method used in G08 and in much of the literature is, \textit{how much is the fit to the data improved by assuming a cross-correlation of the shape predicted by the ISW effect in a \LCDM~model? }   By allowing a free amplitude for the expected cross-correlation shape, G08 found $\Delta \chi^2 = 19$ between the best fit amplitude and the hypothesis of no ISW cross-correlation, and the best amplitude suggested by the data was very close to that predicted by \LCDM.  

Another possible approach is simply to ask whether the null hypothesis, that there is \textit{no}  cross-correlation, has been ruled out, without assuming any particular alternative model. In G08 it was found that a simple chi-square test is still passed, i.e. the null hypothesis is not rejected by current data, having $\chi^2_0 = 67$ for 73 d.o.f.   The absolute chi-square statistic is sensitive to the estimation of the error bars and correlations between measurements; further, ignoring the improvements to the fits when a well-motivated (or even better motivated, given the other evidence) alternative model is considered seems overly pessimistic.   

We have checked that the above-mentioned results by G08 remain similar in the current updated version of the data: 
we obtain for the null hypothesis that $\chi^2_0 = 60.2 $, with 78 d.o.f. For the WMAP7 cosmology this is reduced to $\chi^2_{\mathrm{WMAP}} = 41.6$, and for our best fit amplitude model to $\chi^2_{\mathrm{best}} = 39.7$. At face value, these numbers still do not reject the null hypothesis. Further, the reduced $\chi^2$ is small, potentially indicating that the covariance has been over-estimated. Nonetheless, as mentioned above, the most important quantity is the differential $\Delta \chi^2 $, which indicates a strong preference for the \LCDM~paradigm.

We have also recalculated these results using the TH covariance: in this case we obtain results more in line with the expectations for this number of
degrees of freedom: for the null hypothesis $\chi^2_0 = 88.4 $, which is weakly disfavoured at the 80\% level, and $\chi^2_{\mathrm{WMAP}} = 72.9$. While the difference between TH and MC results is unclear, it is important to notice that the differential between the null hypothesis and \LCDM~remain consistent at  $15 < \Delta \chi^2 < 20$.

\subsubsection {Predictions for best possible ISW measurement} \label{sec:FandP}

While sometimes cited as a skeptical paper regarding the ISW, the work
by \citet{2010MNRAS.406....2F} (FP) does not contradict the earlier ISW literature, though it is perhaps overly conservative in its discussion of the prospects for measuring the  ISW signal. In their study, they repeat the 2MASS cross-correlation
measurement, taking advantage of improved redshift determination of the
2MASS sources to divide the sample in three redshift bins. Their
measurement of the CMB cross-correlation shows no detection, but a weak
preference for \LCDM~models compared to no ISW signal. This is
consistent with most measurements using 2MASS \citep{Rassat:2006kq,Ho:2008bz,Giannantonio:2008zi} apart from the initial claims by \citet{Afshordi:2003xu}. The lack of detection is not surprising because most 2MASS objects are at lower redshifts compared to where the ISW signal is expected to arise. 

However, FP continue to say that the ISW signal might avoid
detection in 10\% of cases, even given ``the best possible data''.  
This seems more pessimistic than the well-known limits on the ISW detection, where the maximum signal-to-noise is capped at 
$(S/N) < 8-10$ for the fiducial \LCDM~cosmology.   
However, this depends significantly on FP's definition of ``the best possible data''; their 10\% prediction assumes a maximum redshift of
$z_{\max} = 0.7$, which effectively cuts out 50\% of the ISW signal, and so is not the best possible data from an ISW perspective.
Including the whole redshift range reduces the number of cases where the ISW effect is not detected by an order of magnitude. 
It is also worth mentioning that FP calculate the fraction of cases where there exists strong evidence for the detection as $\Delta \chi^2 \ge
5$; however, any positive value of $\Delta \chi^2$ would indicate a preference for \LCDM~models compared to no correlation. 

In a companion paper, \citet{2010MNRAS.406...14F} calculate the expected ISW
signal based on the observed 2MASS data. This will be a useful
technique when applied to surveys probing redshifts where the ISW is
most sensitive, as it provides a template to search for in the CMB maps.
For the case of 2MASS, where no cross-correlation is seen, it predicts a
larger ISW signal than is seen in WMAP, which is quite surprising given
the 2MASS could only be sensitive to a fraction of the ISW present.
However, this is primarily due to the much larger than expected power in
their highest redshift bin, which has significantly lower galaxy density
and is more prone to contamination. 

\subsubsection {Significance and field to field fluctuations} \label{sec:LC}

In another paper, \citet{2010A&A...513A...3L} argue that the significances of ISW
detections claimed by many authors are incorrect, based on a
mis-estimation of what they call ``field-to-field fluctuations''. Their
analysis yields comparable results for the CMB-galaxy cross-correlation
function to previous detections, but their estimates for the noise in
the measurements are significantly higher. G08 and independently
\citet{2006MNRAS.372L..23C} estimate these uncertainties around $\simeq 0.2$~
$\mu$K, while \citet{2010A&A...513A...3L} have $\simeq 0.35$ to $0.6
$~$\mu$K.

The reason for this disagreement is unclear. The origin and calculation
methods appear to be comparable to previous approaches; they briefly
discuss jack-knife, Monte Carlo, rotations and analytic methods. For
jack-knives, they find comparable answers to previous approaches, while
their application of other methods yields much higher noise estimates
and they argue that these are more appropriate. We have discussed above
the rotation tests, which we view as consistent with other estimates.
Analytic approaches have been performed before \citep{2007MNRAS.381.1347C}, yielding
results comparable to the other covariance methods, and the explanation of the lack of
agreement with \citet{2010A&A...513A...3L} is not clear. 

Their differences in the Monte Carlo and analytic results are perhaps
hardest to understand; one issue could be that the number of Monte
Carlos that they perform (100) is small compared to what is required for
many purposes. Earlier calculations typically have used thousands of
simulations, in order to ensure convergence in the off-diagonal
covariance and invertibility of the covariance matrix. However, fewer Monte Carlos 
on its own seems unlikely to be able to explain such a large magnitude
difference in the diagonal covariance.

It is worth noting that a subset of the authors of \citet{2010A&A...513A...3L} report a similar inconsistency with estimates of the noise
in the galaxy auto-correlation, drawing into question  previous BAO measurements \citep{2009A&A...505..981S}. 
This suggests the origin for the cross-correlation discrepancy may relate to their estimation of the large-scale structure fluctuations. 
However, to this point the discrepancy has not been explained satisfactorily in either the ISW or BAO context.     

\subsubsection {Comparison to the expected ISW signal}  \label{sec:HM}

\citet{2010A&A...520A.101H} presents an analysis critical of the
ISW detections with the NVSS data set, one of the most analysed data
sets in this context \citep{Boughn:2003yz,Nolta:2003uy,Pietrobon:2006gh,2008MNRAS.384.1289M,Raccanelli:2008bc,2012arXiv1203.3277S}.
His results are somewhat mixed; on the one hand he confirms the
measurement of a signal seen in the cross-correlation, and in the
cross power spectrum ($\ell \sim 10-25 $) at a weaker 2-$\sigma$ level.
However, he claims the theoretical signal should be 5-$\sigma$, significantly larger than what is observed. 
This statement seems at odds with other measurements using this data, which detected the ISW at the expected level but 
with a lower significance.  It is not clear whether the expected amplitude or the inferred errors are responsible for 
this difference.  

\citet{2010A&A...520A.101H} also argues that the correlations arise on smaller scales than
expected for the ISW. In the $\ell \sim 2-10$ region, which he argues should have
half of the ISW signal-to-noise, he finds a low value compared to \citet{Ho:2008bz}, leading him to speculate that an unknown foreground
systematic may be contributing to the observed ISW signal.  Unfortunately, the disagreement with \citet{Ho:2008bz} is not explained, and it is difficult to 
guess its origin given limited details of the methods. 

Hern\'{a}ndez-Monteagudo's statement that half the signal should appear at $\ell < 10$ is somewhat misleading. It is $(S/N)^2$ which adds
cumulatively in multipole, not $S/N$, and only a quarter of the $(S/N)^2 $ is expected for $\ell < 10$. It is true that this implies $S/N$
reaches half of its full value in this range, but this would be equally true for three other independent ranges of $\ell$.  
Unfortunately, it is problematic to take a low significance result and attempt to break it up into subsets where the expected
signal-to-noise is of order one. Estimates of significance are usually taken into account using template fitting techniques which optimally
combine the signal on all scales. 

\subsubsection {Higher statistical significance} \label{sec:higher}

Most detections of the ISW are performed using two-point statistics in
real or multipole space, and they largely yield comparable answers. One exception to this is the main galaxy survey detection found by \citet{2006MNRAS.372L..23C}, which found a higher significance detection (total $S/N = 4.7$ with data from SDSS DR4 alone).
This discrepancy was traced by G08 to be due to a particular cut made by \citet{2006MNRAS.372L..23C}
 on the data (they were excluding galaxies with a large
error on their Petrosian $r$ magnitude); G08 was not able to justify this cut, and to be
conservative used the lower significance answer found without it.
The dependence of the answer on this particular cut
is somewhat worrying and should be folded into the estimate of the systematic uncertainty. However it should be noted that many other similar cuts have been explored,
rarely changing the answers significantly.

In addition, other methods have been used to search for the ISW effect,
including wavelet and stacking methods, and these have sometimes
produced much higher $S/N$ than is seen using the simpler two-point
methods. For example, a number of wavelet methods have been used to analyse the
NVSS cross-correlation  \citep{Vielva:2004zg, 2008MNRAS.384.1289M, Pietrobon:2006gh} and have
sometimes reported higher significances. However, interpretation of the
associated significances are not as straightforward, and often they
explore a large space of possible wavelet shapes and scales and report
the highest significance detection without reference to the expected
theoretical dependence. In such cases, there is an \textit{a posteriori} bias in
the statistics; if one focuses on the cosmological constraints from
all the measurements, these appear to give comparable constraints
on cosmological parameters to those from the two-point statistics. 

Using a different approach, \citet{2008ApJ...683L..99G} have looked for supercluster and supervoids in
the SDSS data; stacking the CMB fields associated with these, they have
observed a temperature hot spot at the positions of the superclusters and
a cold spot associated with the supervoids. The significance claimed is
$4.4 \, \sigma$, much higher than seen in the two-point statistics. Were the
fluctuations Gaussian, which is expected on these large scales, we would
expect that correlating peaks is not as optimal compared to a full
two-point measurement, so the higher significance is surprising.
However, there is some \textit{a posteriori} bias imparted when choosing how many
superclusters to stack, and how large a patch to consider, which may
contribute to the higher significance.

In a later paper \citep{Granett:2008dz} the same authors generated a linear map of the time derivative of the gravitational potential traced by SDSS LRGs, using a Voronoi tessellation technique. While cross-correlations of this map with the CMB reproduced the expected ISW signal at the expected significance level of $\sim 2 \, \sigma$, this ISW map failed to show any signature associated with the superclusters and supervoids of the earlier detection: the mean temperature of the clusters and voids on the ISW map was not significant, as the temperature difference between clusters and voids was found to be compatible with zero, contradicting the suggestion that the signal is due to the linear ISW effect.

The excess signal found in \citet{2008ApJ...683L..99G} has been investigated by various authors \citep{Papai:2010eu,Papai:2010gd,Inoue:2010rp,2011arXiv1109.4126N}, who have shown that it is not easy to explain. Their tests have explored a range of different choices of density profiles for super-voids and -clusters and focused on the linear ISW theory as the origin; however no consensus has been reached on the level of disagreement.
If this excess signal were to be confirmed and shown to be evidence of a higher-than-expected ISW effect, it could suggest a significantly different cosmological model; however, given the novelty of the method and possible new systematics, it is too early to draw any strong conclusions.

\subsection {New galaxy data sets} \label{sec:S09data}

In the coming years, deeper and wider cosmological surveys will improve the significance of ISW measurements (e.g. DES, PanSTARRS, WISE, LSST, Euclid). In the meantime, existing maps can grow to cover more area and/or be re-analysed to obtain new photometric subsamples. For example,  S10 explored three new LRG catalogues at low, medium and high redshifts using the existing SDSS imaging data.  The low redshift catalogue has a depth of $z = 0.35 \pm 0.06$,  similar to the SDSS main galaxy sample, but has relatively few galaxies making Poisson noise a dominant source of error.  The medium redshift 2SLAQ sample most closely approximates the MegaZ LRG sample, but with approximately half as many sources; it has a cross-correlation signal consistent with that seen for the MegaZ data as discussed above.  The most interesting ISW result of S10 arises from their high redshift catalogue based on photometric LRGs calibrated using a pilot AAOmega survey, extending to $z = 0.7$. While the low redshift catalogues simply see lower ISW significance than seen in other analyses, the high--z AAOmega sample appears inconsistent with any ISW detection at all. 

One concern is that the AAOmega sample is pushing the SDSS photometry to its usable limit. For example, AAOmega selected objects with $i$-band de Vaucouleurs magnitudes $ 19.8 < i_{\mathrm{deV}} < 20.5 $, significantly deeper than the cut used by SDSS-III BOSS \citep{Eis11} and studied in \citet{2011MNRAS.417.1350R} of approximately $ i_{\mathrm{deV}} < 19.9$. Furthermore, S10 selected their high-redshift LRGs using the $riz$ colour selection technique as defined by Eqs. (2) through (6) of \citet{Ross:2007qh}. As shown in Fig. 1 of \citet{Ross:2007qh}, these colour cuts can be quite narrow e.g., their priority B objects are defined to be within a colour range of only $0.2 \le (i - z) \le 0.6$, which contains most of the $0.6 < z < 0.7$ AAOmega redshifts. At such faint model magnitudes, we can expect large photometric uncertainties on the observed colours, as demonstrated in Table 6 of \citet{2005astro.ph..8564S}, who found that at $20 < r < 21$ the true (independent) colour errors for red galaxies (like LRGs)  is $0.16$ for $(i-z)$, i.e., nearly half the width of the AAOmega colour selection discussed above.   

We would expect such photometric errors to cause considerable scatter about these AAOmega colour selection boundaries, preferentially leading to contamination from lower redshift and/or lower luminosity interlopers, as there are many more such galaxies in the SDSS sample. Such contamination would lower the correlation function of the AAOmega sample and, interestingly, AAOmega high-redshift sample does possess the lowest auto-correlation function of all three samples used by S10. Finally, it must be remembered that the photometric cuts are based on only 1270 calibration AAOmega spectra, 587 of which were confirmed to be LRGs \citep[see][]{Ross:2007qh}, and this small sample was then extrapolated to 800,000 galaxies in the high-z photometric sample ($\sim 2000$ additional LRG spectra were used to constrain the AAOmega redshift distribution). For comparison, the MegaZ sample we use herein was calibrated using 13,000 spectra.

Stellar contamination is another acknowledged issue. In S10, this contaminant is estimated to be at the $16 \%$ level in the full sample of 800 000 high--z LRGs, and at the $9 \%$ level in a smaller cleaner data set, as also confirmed by the analysis of the auto-correlation of the LRGs \citep{2011MNRAS.416.3033S}. However, even this reduced level is still a significant concern and it could undermine the usefulness of this catalogue for ISW measurements where the expected signal is weak. For comparison, the MegaZ LRG and the SDSS QSO samples used in G08 had both a $ < 5 \%$ stellar contamination, and there it was found that any higher contamination was degrading the data quality  beyond usability for the cross-correlation purposes. While it is expected that a component of random stars ought to have null correlation with the CMB, it may correlate with foregrounds (e.g., Galactic dust and stars both trace the structure of the Galaxy), and it will in any case add to the overall noise level, which is already  high in these studies. Further, the auto-correlation of the AAOmega sample is shown by \citet{2011MNRAS.416.3033S} to decrease significantly on large scales when areas with $A_r > 0.1$ are masked, thus showing that there is a correlation between the AAOmega sample and Galactic foregrounds.

These considerations suggest that the AAOmega sample is not well-suited for an ISW measurement. It is interesting to see whether or not the ISW effect can be detected with this data set, but perhaps it should not be interpreted on the same footing as the other, lower redshift SDSS samples. The lack of a cross-correlation found by S10 should be tracked in the future with better quality data at a similar redshift, but at present it is hard to draw conclusions from it on the general nature of the ISW effect.

\subsection {Systematics}

\subsubsection{The rotations as a test of systematics}

The rotation tests were discussed in detail above; however, we emphasise that systematic contamination would most likely appear associated with rotations about a single axis, and we find no axis with any particular signal beyond what is expected from the estimated covariance.  In general, if the number of outliers were much higher than expected, this would most likely indicate a  mis-estimation of the covariance rather than indicating a particular new systematic contaminant.  The observed consistency seems to confirm our estimates of the significance of the ISW effect.

\subsubsection{Large-scale power in NVSS and bias modelling} \label{sec:spike}

While systematic errors are not pointed to by the rotations, that is not to say there are no causes for concern.  One significant issue with NVSS, highlighted by Hern\'{a}ndez-Monteagudo and in other work \citep{Blake:2001bg,Blake:2004dg}, is that the
auto-correlation signal of NVSS on large scales is significantly higher than what is expected in \LCDM~models, even when considering only the
brightest sources in the NVSS survey. This confirms earlier results on the NVSS \citep{Raccanelli:2008bc} and may imply the existence of a systematic or incorrect modelling of the redshift distribution.  (However, no indications of this have been seen in cross-correlations with other surveys.)   It could also be a true physical effect arising from an evolving or a scale-dependent bias, the latter predicted
by models of primordial non-Gaussianity \citep{Dalal:2007cu, Slosar:2008hx, Afshordi:2008ru, 2008ApJ...677L..77M, 2010PhRvD..81f3530G, Xia:2010yu,Xia:2011hj}. 

Generally, cross-correlations should be more robust to foreground
contaminants than auto-correlation measurements; in this case the radio foregrounds would
have to be correlated with the CMB.  Though not impossible given that radio
sources can emit in the CMB frequencies, it would be surprising if large-scale cross-correlations were introduced at precisely the level to
match cross-correlations seen in other surveys and what is expected
theoretically from \LCDM~models. 

To further test the possible consequences of the excess power in the ACFs, we have modelled this as an extra low-redshift contamination in the catalogues. We have found that we can reproduce the observed excess large-scale auto-correlations by adding a low-redshift Gaussian spike in the
redshift distribution
of the form
\be
\tilde \varphi(z) = \varphi(z) + A_z \cdot \max\left[\varphi(z) \right] \cdot \exp{\left[ - \frac{\left(z - \mu_z \right)^2}{2 \, \sigma_z^2} \right]} \, ,
\ee
where $\max\left[\varphi(z) \right] $ represents the maximum of the unaltered distribution $\varphi(z)$.
Some freedom exists with respect to the parameters of the spike $(A_z, \mu_z, \sigma_z)$, which can also be different for each galaxy catalogue; we have found that a conservative combination for these parameters is $ \mu_z = 0.02, \sigma_z = 0.01$, and $A_z = (0.5, 1, 0.2, 2, 1, 1)$ for the six galaxy catalogues respectively. This is conservative in the sense that it produces more excess power than it is observed.
When using these modified redshift distributions, the CCFs remain unaffected, as the ISW effect has a negligible contribution at these low redshifts; on the other hand, the ACFs increase as expected. When using these spectra for the analytical covariance, we then obtain larger error bars on the cross-correlation due to this extra power. For this particular setting, the total ISW significance drops to $\sim 3.4 \, \sigma$.

This is a rather extreme case, since as the low-redshift spike is present in all catalogues, it will produce high  correlations between the catalogues, which are at odds with the observed density-density correlation functions. To consider an intermediate, more realistic case, we study the scenario where low-$z$ spikes are added where there is a clear excess in the auto-correlations: main galaxies, NVSS, and quasars only, with $A_z = (0, 0.8, 0, 1.5, 0, 1)$. In this case we find that the total significance is $S/N = 3.8$.

Therefore we estimate that even in the worst-case scenario, low-redshift contamination can not affect the significance of our measurement by more than 1 $\sigma$, while in more realistic cases this is significantly reduced, and its effect is therefore compatible with the systematic error of $\pm 0.4 \, \sigma$ which we quote.

\section{Conclusions} \label {sec:concl}

In this paper we have updated our compilation of ISW measurements to the latest available data, finding consistent results with previous studies, a mild (1-$\sigma$) excess signal with respect to the expectations from the \LCDM~model, and an updated overall significance of $ S/N = 4.4 \pm 0.4 $. We have performed  additional tests on the combined
ISW measurements, finding no evidence for systematics. In particular, we concluded that the rotation test is not an issue for our data, and appears to be of little significance for the data by S10.
We have shown that our correlation data remain robust with the latest WMAP7 release of CMB data, and with the final SDSS DR8 imaging. 
We have discussed the impact of several issues and criticisms which have arisen in the literature in the recent years, concluding that most of the issues are not serious problems for the use of the ISW as a cosmological probe.
We have publicly released all the data, including the maps and the masks for the LSS catalogues.
Improvements in the statistical analysis and cosmological consequences will be presented in a forthcoming paper.

It is clear that, if the \LCDM~model is the true underlying model of cosmology, the significance of the ISW effect will remain lower than some other cosmological probes; however, it represents nonetheless a unique signal which allows us to independently confirm the presence of dark energy through its impact on structure growth and potentially detect deviations in how gravity works to build cosmic structures. %
Upcoming data from ongoing and future surveys, such as e.g. DES, PanSTARRS, WISE, LSST and Euclid, will be crucial for answering these questions, and to push the significance of the ISW detections close to its theoretical limits.

\section*{Acknowledgements}
We thank Stephen Boughn, Martin Kilbinger, Will Percival, Stefanie Phleps, and Alvise Raccanelli for useful discussions.
TG acknowledges support from the Alexander von Humboldt Foundation, and from the Trans-Regional Collaborative Research Centre TRR 33 -- ``The Dark Universe''. RC, BN and AR gratefully acknowledge financial support from the STFC via the rolling grant ST/I001204/1. 

Funding for SDSS-III has been provided by the Alfred P. Sloan Foundation, the Participating Institutions, the National Science Foundation, and the U.S. Department of Energy Office of Science. The SDSS-III web site is http://www.sdss3.org/.

SDSS-III is managed by the Astrophysical Research Consortium for the Participating Institutions of the SDSS-III Collaboration including the University of Arizona, the Brazilian Participation Group, Brookhaven National Laboratory, University of Cambridge, Carnegie Mellon University, University of Florida, the French Participation Group, the German Participation Group, Harvard University, the Instituto de Astrofisica de Canarias, the Michigan State/Notre Dame/JINA Participation Group, Johns Hopkins University, Lawrence Berkeley National Laboratory, Max Planck Institute for Astrophysics, Max Planck Institute for Extraterrestrial Physics, New Mexico State University, New York University, Ohio State University, Pennsylvania State University, University of Portsmouth, Princeton University, the Spanish Participation Group, University of Tokyo, University of Utah, Vanderbilt University, University of Virginia, University of Washington, and Yale University.
 \bibliographystyle{mn2e}
 \bibliography{ms}

\label{lastpage}

\end{document}